\documentclass{JHEP3}
\usepackage{epsfig}
\usepackage{epstopdf}
\input{epsf.sty}
\usepackage{epsf}
\usepackage{amssymb}
\def\be{\begin{equation}}
\def\ee{\end{equation}}
\def\ba{\begin{eqnarray}}
\def\ea{\end{eqnarray}}
\include{mydefs}

\title{Signatures of Primordial non-Gaussianities in the Matter Power-Spectrum and Bispectrum:\\ the Time-RG Approach}

\author{Nicola Bartolo$^{1,2}$, Juan P. Beltr\'an Almeida$^{3}$, Sabino Matarrese$^{1,2}$,  Massimo Pietroni$^{2}$, and Antonio Riotto$^{2,4}$ \\

$^1$ Dipartimento di Fisica ``G. Galilei'', Universit\`{a} degli Studi di 
Padova, \\ via Marzolo 8, I-35131 Padova, Italy
\vskip 0.3cm
$^2$ INFN, Sezione di Padova, via Marzolo 8, I-35131 Padova, Italy
\vskip 0.3cm
$^3$  Instituto de F\'{\i}sica Te\'orica, UNESP - Universidade Estadual Paulista, \\ Caixa Postal 70532-2, 01156-970 S\~ao Paulo, SP, Brazil
\vskip 0.3cm
$^4$ CERN, Theory Division, CH-1211 Geneva 23, Switzerland

\vskip 0.3cm
Emails: nicola.bartolo@pd.infn.it, jalmeida@ift.unesp.br, sabino.matarrese@pd.infn.it, massimo.pietroni@pd.infn.it  and riotto@mail.cern.ch
\vskip 0.3cm
 CERN-PH-TH/2009-260}

\abstract{ 
We apply the time-renormalization group  approach  to study the effect of primordial non-Gaussianities in the non-linear 
evolution of cosmological dark matter  density perturbations. This method  improves the standard perturbation approach  by solving renormalization group-like  equations governing 
the dynamics of gravitational instability. The  primordial bispectra constructed from the dark matter  density contrast and the velocity fields represent   initial conditions for the renormalization 
group flow. We consider local, equilateral and folded 
shapes  for the initial non-Gaussianity  and analyze  as well the case in which the non-linear parameter
$f_{\rm NL}$ parametrizing the strength of the non-Gaussianity depends on the momenta
in Fourier space through a power-law relation, the so-called running non-Gaussianity.
For the local model of non-Gaussianity we compare our findings for the 
power-spectrum with those of recent N-body simulations and find that they accurately fit  the N-body data up to 
wave-numbers $k \sim 0.25\, h$/Mpc at $z=0$. We also present predictions for the (reduced) matter bispectra for the various shapes of non-Gaussianity. 

}


\begin{document}

\maketitle

\section{Introduction}

During the last years, the use of semi-analytic methods to study the Large Scale Structure (LSS) formation 
via gravitational instability has experienced a renewed interest, 
motivated mainly by high precision measurements of statistical properties expected in the next generation of galaxy surveys. 
These methods typically involve resummation or closure prescriptions and need to be  compared with numerical simulations to assess their range of validity.
Renomalized perturbation theory   ~\cite{Crocce:2005xy,Crocce:2005xz} is based on suitable resummations of perturbative terms and is expected to work well for the linear 
and ``weakly non-linear" regime.
Perturbation theory can be 
rewritten in a compact way which makes it possible the use of standard tools of quantum field theory, thus allowing a systematic treatment of perturbative terms  in terms only of a finite set of basic building blocks, namely, the initial fields, 
the {\it linear propagator}, which describes the linear evolution of the initial fields and an {\it interaction vertex}, responsible for the non linear mode 
coupling of the fields.    This leads to the resummation of infinite classes of perturbation 
theory diagrams. The renormalization group perturbation theory \cite{mc} attempts to regulate the relative divergence of one-loop standard perturbation theory (for a review, see \cite{reviewsc}) using renormalization
group methods. A path-integral formulation of the Vlasov equation has been developed in \cite{vale}
in terms of the distribution function in the phase space. The lagrangian resummation theory
\cite{mat} is an extension of the well-developed Lagrangian perturbation theory. 
Refs.~\cite{Matarrese:2007aj} and \cite{Matarrese:2007wc}  proposed  a formalism based on 
path-integrals and renormalization group techniques which provides a systematic resummation scheme of cosmological perturbation theory (subsequently, this formalism was extended in Ref.~\cite{Izumi:2007su} to calculate the non-linear propagator in the presence of non-Gaussian initial conditions of the fields).
Finally, in the Time-Renormalization Group (TRG) approach proposed in Ref.~\cite{Pietroni:2008jx}, the power spectrum, the bispectrum and higher order correlations, are obtained -- at any redshift and for any momentum scale -- by integrating a system of differential equations. The method is similar to the familiar BBGKY hierarchy. Truncating at the level of the trispectrum, the solution of the equations corresponds to the summation of an infinite class of perturbative corrections. 
The approach can be seen as a particular formulation of the renormalization group, in which time is the flow parameter.
Compared to other resummation frameworks, this
 scheme  is particularly suited to cosmologies other than $\Lambda$CDM and has been recently applied to
 compute the non-linear spectrum in presence of neutrino masses \cite{nu}.
A critical look at the various cosmological perturbation theory techniques can be found in \cite{martinw}.

On the other hand, over the last
decade a great deal of	evidence has been accumulated from the Cosmic Microwave
Background (CMB) anisotropy and Large Scale Structure (LSS) spectra
that the observed
structures originated from seed fluctuations generated during a primordial
stage of inflation. While standard single-field models of
slow-roll inflation
predict that these fluctuations are very close to  
Gaussian (see \cite{acquaviva,maldacena}), 
non-standard scenarios allow for a larger level of non-Gaussianity (NG)
(see \cite{bartoloreview} 
and references therein).  A signal is  gaussian if the information it carries
is completely
encoded in the two-point correlation function, all higher connected correlators
being
zero.
Deviations from Gaussianity are therefore  encoded, {\it e.g.}, in the connected 
 three- and four-point correlation functions which are dubbed the bispectrum
and the trispectrum,
respectively. A phenomenological way of parametrizing the level of NG is to
expand the fully 
 non-linear primordial Bardeen
 gravitational potential $\Phi$ in powers of the linear gravitational potential
$\Phi_{\rm L}$
 
 \be
 \label{phi}
 \Phi=\Phi_{\rm L}+f_{\rm NL}\left(\Phi_{\rm L}^2-\langle\Phi_{\rm
L}^2\rangle\right)\, .
 \ee
The  dimensionless
quantity  $f_{\rm NL}$   sets the
magnitude of the three-point 
correlation function \cite{bartoloreview}. If the process generating the primordial
NG  
is local in space, the parameter  $f_{\rm NL}$ in Fourier
space is
independent of the momenta
entering the corresponding  correlation functions; if instead the process which
generates the
primordial cosmological perturbations is
non-local in configuration space, like in 
models of inflation with non-canonical kinetic terms, $f_{\rm NL}$  
acquires a dependence on the momenta.  
It is clear that  detecting a significant amount of 
NG and its shape either from the CMB or from the
LSS offers the possibility of opening a   window  into the 
dynamics of the universe
during the very first 
stages of its
evolution. Non-Gaussianities are particularly relevant in the  high-mass end of
the power spectrum of perturbations, {\it i.e.} on the scale of galaxy clusters,
since the effect of non-Gaussian fluctuations becomes especially visible on
the  tail of the probability distribution. 
As a result, both 
the abundance and  the clustering properties of very massive halos
are sensitive probes of primordial 
non-Gaussianities \cite{MLB,GW,LMV,MMLM,KOYAMA,MVJ,RB,RGS,bi, tri}, 
and could be detected or significantly constrained by
the various planned large-scale galaxy surveys,
both ground based (such as DES, PanSTARRS and LSST) and on satellite
(such as EUCLID and ADEPT) see,  e.g.  \cite{ks} and \cite{CVM},~\cite{MV2009}. 
Furthermore, the primordial NG
alters the clustering of dark matter (DM) halos inducing a scale-dependent
bias on large 
scales \cite{Dalal,MV,slosar,tolley}  while even for small primordial
NG the evolution of perturbations on super-Hubble scales yields
extra
contributions on 
smaller scales \cite{bartolosig,MV2009}.  
The strongest current
limits on the strength of local NG set the $f_{\rm NL}$
parameter to be in the range $-4<f_{\rm NL}<80$ at 95\% confidence level
\cite{zal}. 

In this paper, we implement the TRG approach  to investigate the effects imprinted by a  primordial 
NG  in  the non-linear evolution of cosmological perturbations. In particular, we will compute
the DM power spectrum and the (reduced) bispectrum when some primordial initial
NG condition is present. In the TRG approach this information is  promptly encoded in the RG equations
for the power spectrum and the bispectrum. As a consequence, we can easily study the impact of the
various shapes of NG. In particular, we consider the local, the equilateral and the folded shapes. 
As we shall see, our approach is based on a closure assumption, {\it i.e.}
we solve the RG equations for the power spectrum and the bispectrum, setting the connected
$n$-point correlators to zero starting from the four-point correlator, the trispectrum. 
While  this assumption  represents a clear limitation of our  approach, nevetheless  it  provides a 
well-defined and controllable  
scheme which allows us to quantify in a precise form the theoretical error associated with the method. 
For the local model of non-Gaussianity we will  compare our findings for the 
power-spectrum with those of recent N-body simulations and find that they accurately fit  
rather well the N-body data up to 
wave-numbers $k \sim 0.25\, h$/Mpc. 
We will also present predictions for the (reduced) matter bispectra for the various shapes of NG.

The paper is organized as follows. In Section 2 we review the time evolution of the correlators governed by Eulerian dynamics and discuss our truncation hypothesis in comparison with other approaches
in the literature. 
In Section 3 we consider the primordial non-Gaussianities compatible with the truncation scheme of this paper and some 
theoretically motivated models for the primordial bispectrum. In Section 4, we introduce the reduced bispectrum as a relevant tool to 
study the effects related to the shape of the primordial bispectrum. In Section 5 we present our results for the power-spectrum and a comparison 
with the results of N-body simulations for ``local'' NG. We also present several illustrative plots of the reduced 
bispectrum for each model of NG. In Section 6, we end with our conclusions and discuss future perspectives on this line of work.

\section{Dynamics of Gravitational Instability}
\subsection{Eulerian Dynamics}
Our starting point are the hydrodynamic equations in the ``single stream" approximation of a self-gravitating fluid made of cold dark matter 
collisionless particles in an expanding Universe. In terms of the mass-density fluctuation $\delta$, the  peculiar velocity ${\bf v}$ of the 
fluid and the peculiar gravitational potential $\phi$, the dynamics of the system is governed by the system of equations 
\ba\label{eomx}
&&\frac{\partial \delta}{\partial \tau} + \nabla \cdot[(1+\delta){\bf{v}}] = 0,
{}  \\ 
&&\frac{\partial {\bf{v}}}{\partial \tau} + {\cal{H}}{\bf{v}} + ({\bf{v}}\cdot \nabla){\bf{v}} = -\nabla \phi,
{}  \\ 
&&\nabla^2\phi = \frac{3}{2}{\cal{H}}^2 \Omega_{m}\delta.
\ea
The first two equations are the continuity and Euler equations respectively, while the third one is the Poisson equation obeyed by 
perturbations on sub-horizon scales. Here, $a$ is the scale factor of the background, $\tau = \int dt /a$ is the conformal time, 
${\cal{H}} \equiv d\log a/d\tau = a H$ and $\Omega_{m}$ is the matter density parameter. In the following we restrict our discussion 
to an Einstein-de Sitter model, so we take $\Omega_{m}=1$\footnote {For a generalization to more general cosmological backgrounds 
the reader is referred to Ref.~\cite{Pietroni:2008jx}.}.  As usual, we take the divergence of the Euler equation and define the velocity divergence 
$\theta = \nabla \cdot {\bf v}$ and Fourier transform. The resulting equations can be written in a compact notation as \cite{Crocce:2005xy}
\be\label{eom}
(\delta_{ab}\partial_{\eta} + \Omega_{ab})\varphi_b ({\bf k}, \eta) = e^{\eta}\gamma_{abc}({\bf k, -p, -q})\varphi_b ({\bf p}, \eta)\varphi_c ({\bf q}, \eta),
\ee
where we define the two-component field $\varphi_{a} (a= 1,2)$ through 
\ba\label{phi-dt}
\left( \begin{array}{c} 
\varphi_1 ({\bf k}, \eta) \\
\varphi_2 ({\bf k}, \eta)
\end{array} \right) \equiv
\exp (-\eta)\left( \begin{array}{c} 
\delta ({\bf k}, \eta) \\
-\theta ({\bf k}, \eta)/{\cal H}
\end{array} \right),
\ea
$e^{\eta} = a/a_{in}$ for an initial scale factor $a_{in}$ conveniently fixed at an early epoch,
\ba\label{Omega}
{\bf \Omega} = \left( \begin{array}{cc} 
1 & -1 \\
-3/2 & 3/2
\end{array} \right)
\ea
and $\gamma_{abc}$ is a {\it vertex} function which describes the non-linear coupling 
of the modes and whose only non-vanishing components are
\ba\label{gcomponents}
&&\gamma_{121}({\bf p,\, q,\, k})=\frac{1}{2}\delta ({\bf p+q+k})\frac{{\bf (q+k)\cdot q}}{q^2},\\
&&\gamma_{222}({\bf p,\, q,\, k})=\frac{1}{2}\delta ({\bf p+q+k})\frac{{\bf (q+k)^2 q\cdot k}}{k^2 q^2},\\
&&\gamma_{112}({\bf p,\, q,\, k})=\gamma_{121}({\bf p,\, k,\, q}).
\ea
Summation (integration) over repeated index (momenta) is understood in Eq.~(\ref{eom}).

\subsection{Time evolution of the correlators}
\begin{figure}[h] 
\centering
\includegraphics[scale=.7]{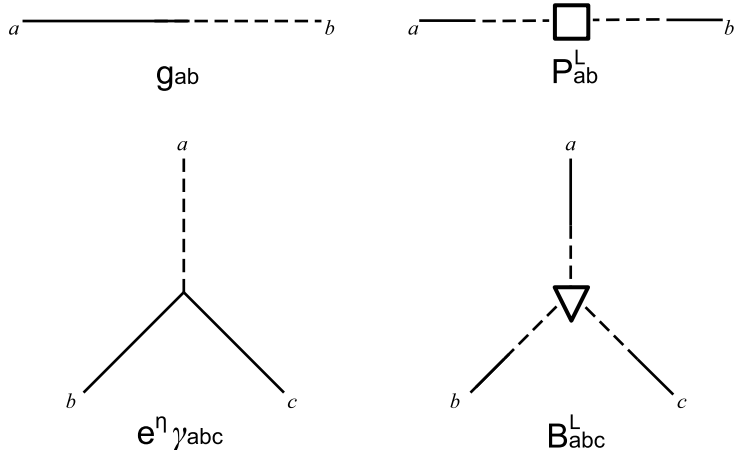}
\caption{Basic building blocks for the perturbative approach with non-gaussian initial conditions. The linear propagator $g_{ab}(\eta_a , \eta_b)$ is causally oriented from the time $\eta_b$ (dashed segment) to $\eta_a$ (solid segment). The interaction vertex $e^{\eta}\gamma_{abc}$ introduces the non linear mode 
coupling of the fields. The square on the linear power spectrum diagram, represents the initial conditions of the power spectrum, while the triangle in the linear bispectrum diagram represents the initial bispectrum, given by the sum of the newtonian contribution at $z=z_{in}$ and the contribution due to primordial non-Gaussianities. }\label{fey-diag}
\end{figure}

\begin{figure}[h] 
\centering
\includegraphics[scale=.7]{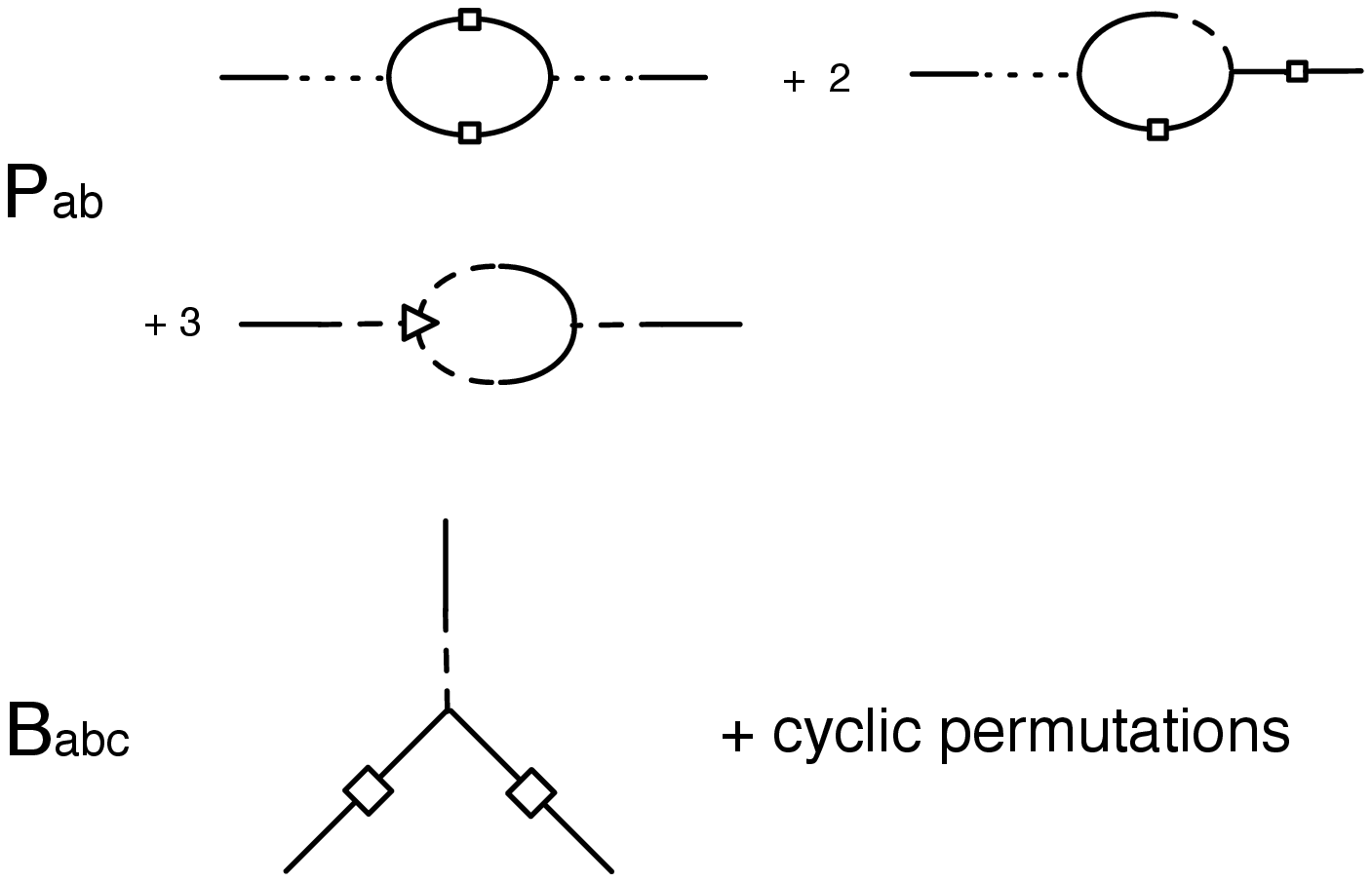}
\caption{Diagrammatic representation of the formal solution for PS and BS. The first two lines represents the lowest order contributions to the PS equation. The first line is the  ${\cal O}(\gamma^2)$ contribution to the PS. The second line is the  contribution coming from the non-vanishing initial BS which is a tree level, ${\cal O}(\gamma)$, contribution. The the third line is the tree level, ${\cal} O(\gamma)$, contribution  to the BS. }\label{one-loop}
\includegraphics[scale=.7]{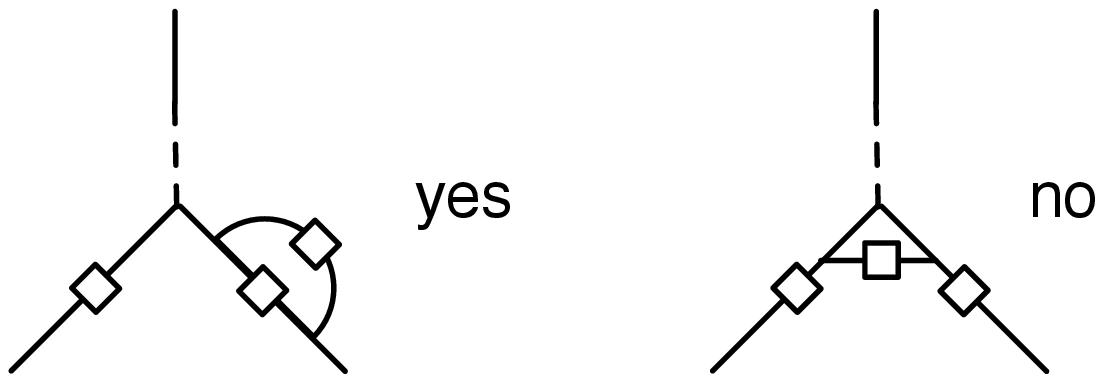} 
\caption{Perturbative corrections to ${\cal O}(\gamma^3)$ order for the bispectrum which are included (left), and not included (right) by implementing the truncation prescription $T_{abcd}=0$. }\label{truncation}

\end{figure}

Following Ref.~\cite{Pietroni:2008jx} the time evolution of the field correlators can be obtained directly from iterative application of the equations of motion
(\ref{eom}). The result is an infinite tower of coupled  differential equations relating the evolution of $n$-point correlators with the $n$ and $(n+1)$-point correlators evaluated at the same time: 
\ba\label{tower}
 \partial_\eta\,\langle \varphi_a \varphi_b\rangle &=& -\Omega_{ac} 
\langle \varphi_c \varphi_b\rangle   - 
\Omega_{bc} 
\langle \varphi_a \varphi_c\rangle 
\nonumber\\
&&+e^\eta \gamma_{acd}\langle \varphi_c\varphi_d \varphi_b\rangle +e^\eta \gamma_{bcd}\langle \varphi_a\varphi_c \varphi_d\rangle\,,\nonumber\\
&&\nonumber\\
 \partial_\eta\,\langle \varphi_a \varphi_b  \varphi_c \rangle & =&  -\Omega_{ad} 
\langle \varphi_d \varphi_b\varphi_c\rangle  -\Omega_{bd} 
\langle \varphi_a \varphi_d\varphi_c\rangle  -\Omega_{cd} 
\langle \varphi_a \varphi_b\varphi_d\rangle
\nonumber\\
&&+e^\eta \gamma_{ade}\langle \varphi_d\varphi_e \varphi_b\varphi_c\rangle  
+e^\eta \gamma_{bde}\langle \varphi_a\varphi_d \varphi_e\varphi_c\rangle \nonumber\\
&&+e^\eta \gamma_{cde}\langle \varphi_a\varphi_b \varphi_d\varphi_e\rangle \,,
\nonumber \\
&&\nonumber\\
 \partial_\eta\,\langle \varphi_a \varphi_b  \varphi_c  \varphi_d \rangle &=& \cdots\nonumber\\
 &&\nonumber\\
& \cdots&
\ea
Here, we omitted the time and momentum dependence in order to have compact expressions. To solve this system of equations is equivalent to apply the standard perturbation theory approach in which one calculates the n-point correlation functions by summing an infinite series of perturbative corrections depending on the interaction vertex $\gamma$, the red-shift and the initial statistic of the fields. Next, we introduce the following nomenclature for  the first correlation functions of the fields
\ba 
&&\langle \varphi_a ({\bf k}, \eta)\varphi_b({\bf q}, \eta)\rangle \equiv  \delta_{D}({\bf k}+{\bf q})P_{ab}({\bf k}, \eta)\nonumber\\ 
&&\langle \varphi_a ({\bf k}, \eta)\varphi_b({\bf q}, \eta) \varphi_c({\bf p}, \eta)\rangle \equiv  \delta_{D}({\bf k}+{\bf q}+{\bf p})B_{abc}({\bf k},\, {\bf q},\, {\bf p};\, \eta)\nonumber\\   
&&\langle \varphi_a ({\bf k}, \eta)\varphi_b({\bf q}, \eta) \varphi_c({\bf p}, \eta)\varphi_d({\bf r}, \eta)\rangle \equiv  \delta_{D}({\bf k}+{\bf q})\delta_{D}({\bf p}+{\bf r})P_{ab}({\bf k}, \eta)P_{cd}({\bf p}, \eta)\nonumber\\ 
&& \qquad +\delta_{D}({\bf k}+{\bf p})\delta_{D}({\bf q}+{\bf r})P_{ac}({\bf k}, \eta)P_{bd}({\bf q}, \eta) +\delta_{D}({\bf k}+{\bf r})\delta_{D}({\bf q}+{\bf p})P_{ad}({\bf k}, \eta)P_{bc}({\bf q}, \eta)\nonumber\\ 
&& \qquad + \delta_{D}({\bf k}+{\bf q}+{\bf p}+{\bf r})T_{abcd}({\bf k},\, {\bf q},\, {\bf p},\, {\bf r};\, \eta)\nonumber ,
\ea
where $P_{ab}$ is the power spectrum (PS) , $B_{abc}$ is the bispectrum (BS), and $T_{abcd}$ is the connected part of the four-point function, the {\it trispectrum}.  
As said before, if we want to calculate the time evolution of the power-spectrum it will involve the time evolution of the bispectrum, 
which in turn involves the trispectrum and so on, giving us the infinite tower of equations (\ref{tower}). 
This system can be truncated if we neglect the trispectrum $T_{abcd}=0$, letting us with the following system 
of two equations for the PS and the BS: 
\ba\label{system}
&& \partial_\eta\,P_{ab}({\bf k}\,,\eta) = - \Omega_{ac} ({\bf k}\,,\eta)P_{cb}({\bf k}\,,\eta)  - \Omega_{bc} ({\bf k}\,,\eta)P_{ac}({\bf k}\,,\eta) \nonumber\\ 
&&\qquad\qquad\quad\quad+e^\eta \int d^3 q\, \left[ \gamma_{acd}({\bf k},\,{\bf -q},\,{\bf q-k})\,B_{bcd}({\bf k},\,{\bf -q},\,{\bf q-k};\,\eta)\right.\nonumber\\ 
&&\qquad\qquad\qquad\qquad\qquad\left. + B_{acd}({\bf k},\,{\bf -q},\,{\bf q-k};\,\eta)\,\gamma_{bcd}({\bf k},\,{\bf -q},\,{\bf q-k})\right]\,,\nonumber\\ 
&&\nonumber\\ 
&&  \partial_\eta\,B_{abc}({\bf k},\,{\bf -q},\,{\bf q-k};\,\eta) =  - \Omega_{ad} ({\bf k}\,,\eta)B_{dbc}({\bf k},\,{\bf -q},\,{\bf q-k};\,\eta)\nonumber\\ 
&&\qquad\qquad\qquad\qquad\qquad\quad- \Omega_{bd} ({\bf -q}\,,\eta)B_{adc}({\bf k},\,{\bf -q},\,{\bf q-k};\,\eta)\nonumber\\ 
&&\qquad\qquad\qquad\qquad\qquad\quad - \Omega_{cd} ({\bf q-k}\,,\eta)B_{abd}({\bf k},\,{\bf -q},\,{\bf q-k};\,\eta)\nonumber\\ 
&&\qquad\qquad\qquad\qquad\qquad\quad + 2 e^\eta \left[ \gamma_{ade}({\bf k},\,{\bf -q},\,{\bf q-k}) P_{db}({\bf q}\,,\eta)P_{ec}({\bf k-q}\,,\eta)\right.\nonumber\\ 
&&\qquad\qquad\qquad\qquad\qquad\quad +\gamma_{bde}({\bf -q},\,{\bf q-k},\,{\bf k}) P_{dc}({\bf k-q}\,,\eta)P_{ea}({\bf k}\,,\eta)\nonumber\\ 
&&\qquad\qquad\qquad\qquad\qquad\quad +\left. \gamma_{cde}({\bf q-k},\,{\bf k},\,{\bf -q}) P_{da}({\bf k}\,,\eta)P_{eb}({\bf q}\,,\eta)\right]\,.
\ea
The system can be formally solved in the form
\ba\label{formals} 
&& P_{ab}({\bf k}\,,\eta) =  g_{ac}(\eta , 0)g_{bd}(\eta , 0)  P_{cd}({\bf k}\,,\eta=0)   \nonumber\\ 
&&\qquad\qquad\quad\quad+\int_0^{\eta}d\eta'e^{\eta'} \int d^3 q\,g_{ae}(\eta , \eta')g_{bf}(\eta , \eta') \nonumber\\ 
&&\qquad\qquad\quad\quad\times \left[ \gamma_{ecd}({\bf k},\,{\bf -q},\,{\bf q-k})\,B_{fcd}({\bf k},\,{\bf -q},\,{\bf q-k};\,\eta') + (e \longleftrightarrow f)\right]\nonumber\\ 
&&  B_{abc}({\bf k},\,{\bf -q},\,{\bf q-k};\,\eta) =   g_{ad}(\eta , 0)g_{be}(\eta , 0) g_{cf}(\eta , 0) B_{def}({\bf k},\,{\bf -q},\,{\bf q-k};\eta=0) \nonumber\\ 
&&\qquad\qquad\qquad\qquad\quad + 2 \int_0^{\eta}d\eta'e^{\eta'} g_{ad}(\eta , \eta')g_{be}(\eta , \eta')g_{cf}(\eta , \eta')\nonumber\\ 
&&\qquad\qquad\qquad\qquad\quad \times\left[ \gamma_{dgh}({\bf k},\,{\bf -q},\,{\bf q-k}) P_{eg}({\bf q}\,,\eta')P_{fh}({\bf k-q}\,,\eta')\right.\nonumber\\ 
&&\qquad\qquad\qquad\qquad\quad +\gamma_{egh}({\bf -q},\,{\bf q-k},\,{\bf k}) P_{fg}({\bf k-q}\,,\eta')P_{dh}({\bf k}\,,\eta')\nonumber\\ 
&&\qquad\qquad\qquad\qquad\quad +\left. \gamma_{fgh}({\bf q-k},\,{\bf k},\,{\bf -q}) P_{dg}({\bf k}\,,\eta')P_{eh}({\bf q}\,,\eta')\right]\,,
\ea
where $g_{ac}({\bf k}, \eta , \eta')$ is the {\it linear propagator} which is the Green's function of the linearized version of equation (\ref{eom}) (by setting $\gamma =0$) and gives the time evolution of the fields at linear order: $\varphi_a^{L}({\bf k}, \eta) = g_{ac}({\bf k}, \eta , \eta')\varphi_c^{L}({\bf k}, \eta')$. The subscript $L$ stands for the linear order approximation. 
From Eq. (\ref{formals}) one gets an insight on the effect of our only approximation, namely,  the truncation prescription $T_{abcd}=0$. First, the solution for the bispectrum is formally ``tree-level", {\it i.e.} it contains no momentum integration. It has the same structure as the lowest perturbative contribution to the bispectrum, but with the linear power spectra replaced by the resummed, time-dependent ones. Inserting the bispectrum in the solution for the power spectrum, we see that it is formally one loop (one momentum integral), but with resummed power spectra in the loops instead of linear ones. In diagrammatic terms, starting from the one loop diagrams for the power spectrum (which have been computed by Taruya et al.  \cite{Taruya:2008pg}, for the 
non-Gaussian case), one is adding all the infinite contributions which can be obtained by adding corrections to the internal power spectra lines at any perturbative order. We use the language of Feynman diagrams introduced in \cite{Matarrese:2007aj, Matarrese:2007wc, Pietroni:2008jx} to represent diagrammatically the perturbative contributions to the correlators. The basic building blocks of this diagrammatic approach are shown in figure \ref{fey-diag}. In terms of them, we can represent the lowest order terms of this procedure as shown in figure \ref{one-loop}. 
On the other hand, the comparison of the bispectrum with the computation by Sefusatti \cite{Sefusatti:2009qh} is not so straightforward. Compared to that computation we, again, add the corrections to the power spectra lines at all orders, but we do not include the corrections to the vertex, which Sefusatti computed at one-loop. In order to do that, we should go beyond our truncation approximation and include the trispectrum in the hierarchy of our equations.  
This can be easily understood if one recalls that in perturbation theory the connected four-point correlator always gets a contribution which is generated by the bispectrum. It is precisely this contribution
which generates the one-loop correction to the bispectrum which can be seen as a correction to the
interaction vertices $\gamma$. Diagrammatically, the terms included and not included in this approximation are shown in figure \ref{truncation}. Including the trispectrum is straightforward in the TRG approach and 
we leave this to future work. On general grounds, we expect that the inclusion of the
trispectrum will push the length scale where the method works to smaller scales. 
For a more detailed explanation of the diagrammatic of the method the reader is referred to Sections 3 and 4 in Ref.~\cite{Pietroni:2008jx}.

\section{Bispectrum Initial Conditions}
In this section we calculate the initial conditions, at redshift $z=z_{in}$, for the system (\ref{system}) for our case of non-Gaussian primordial statistics. To this end,  we calculate the three-point correlators for the matter density contrast and velocity divergence up to second order terms taking into account the effect imprinted by primordial non-Gaussianities. In other words, we assume that second order perturbation theory holds from $z=\infty$ to $z=z_{in}$, and the resummation embodied by the TRG  is active from $z_{in}$ down to $z=0$. We follow the notations and results of Ref.~\cite{bartolosig} where the effects of primordial non-Gaussianities in the general relativistic cosmological evolution of matter 
perturbations were calculated up to second order in both the comoving and Poisson gauge, deep in the matter dominated epoch. \\

Under certain conditions that we shall specify shortly, the results of \cite{bartolosig} can be extended to the case of a $\Lambda$CDM cosmology. In this case, the linear order solution of the fields are $\delta^{(1)}_{{\bf k}} (\tau)= D_{+}(a)\delta^{(0)}_{{\bf k}}$ and $\theta^{(1)}_{{\bf k}} (\tau)= -{\cal H}f(\tau)D_{+}(a)\delta^{(0)}_{{\bf k}}$ where $D_{+}(a)$ is the linear growth factor of density perturbations and $f(\tau)=d \ln D_{+}(a)/d \ln a$. Information about the cosmological parameters of the background will be encoded in $D_{+}(a)$ and $f(\tau)$. At this point, being interested in fixing the initial conditions at some early $z_{in}$, 
we use the approximation  $\Omega_{m}/f(\tau)^2 \approx 1$ which, as discussed 
in Ref.~\cite{reviewsc}(see also Appendix B of Ref.~\cite{Scoc}), is a condition 
respected during most of the cosmological time evolution. Within this approximation one can simply use the kernels as can be derived from Ref.~\cite{bartolosig}, properly replacing the conformal time 
$\tau$ in terms of the functions $D_{+}(a)$ and $f(\tau)$. Then, the second order expansions for the matter density perturbation and velocity divergence acquire the form
\be\label{2ndd}
\delta_{{\bf k}} (\tau)= \delta^{(1)}_{{\bf k}} (\tau)+\frac{1}{2} \delta^{(2)}_{{\bf k}}(\tau)= \delta^{(1)}_{{\bf k}} (\tau)+\int d^3{\bf k}_1d^3{\bf k}_2\,{\cal K}_{\delta}({\bf k}_1,{\bf k}_2;\tau) \delta^{(1)}_{{\bf k}_1}(\tau) \delta^{(1)}_{{\bf k}_2}(\tau)\delta_D ({\bf k}_{12}-{\bf k}),
\ee
\be\label{2ndth}
\theta_{{\bf k}} (\tau)= \theta^{(1)}_{{\bf k}} (\tau)+\frac{1}{2} \theta^{(2)}_{{\bf k}}(\tau) = \theta^{(1)}_{{\bf k}} (\tau)+\int d^3{\bf k}_1d^3{\bf k}_2\,{\cal K}_{\theta}({\bf k}_1,{\bf k}_2;\tau) \theta^{(1)}_{{\bf k}_1}(\tau) \theta^{(1)}_{{\bf k}_2}(\tau)\delta_D ({\bf k}_{12}-{\bf k}),
\ee
where we introduced the notation ${\bf k}_{ij\cdots}={\bf k}_{i}+{\bf k}_{j}+ \cdots$ and, following the procedure and the results in \cite{bartolosig} we find that
\ba\label{kd}
{\cal K}_{\delta}({\bf k}_1,{\bf k}_2;\tau)  &=&  {\cal K}_{\delta}^N({\bf k}_1,{\bf k}_2) +6f_{\rm NL}^{\delta}({\bf k}_1,{\bf k}_2)\frac{k^2 E(\tau)}{ k_1^2 k_2^2},\\
\label{kth}
{\cal K}_{\theta}({\bf k}_1,{\bf k}_2;\tau)  &=& -\frac{1}{{\cal H}f(\tau)}\left[{\cal K}_{\theta}^N({\bf k}_1,{\bf k}_2)+6 f_{\rm NL}^{\theta}({\bf k}_1,{\bf k}_2)\frac{k^2 E(\tau)}{ k_1^2 k_2^2}\right],
\ea
where $E(\tau)\equiv H_0^2 \Omega_{m,0}/(4 D_{+}(a))$. In the equations (\ref{kd})-(\ref{kth}) 
\ba\label{KN}
{\cal K}_{\delta}^N({\bf k}_1,{\bf k}_2)  &\equiv &  \frac{5}{7}+\frac{2}{7}\frac{({\bf k}_1\cdot {\bf k}_2)^2}{k_1^2 k_2^2} + \frac{1}{2}\frac{{\bf k}_1\cdot {\bf k}_2 \left(k_1^2+k_2^2\right)}{k_1^2 k_2^2}\\
{\cal K}_{\theta}^N({\bf k}_1,{\bf k}_2)  &\equiv & \frac{3}{7}+\frac{4}{7}\frac{({\bf k}_1\cdot {\bf k}_2)^2}{k_1^2 k_2^2}+\frac{1}{2}\frac{{\bf k}_1\cdot {\bf k}_2 \left(k_1^2+k_2^2\right)}{k_1^2 k_2^2}
\ea
are the standard one-loop Newtonian kernels for the expansion of the fields \cite{CM}, and 
\ba\label{fnldd}
f_{\rm NL}^{\delta}({\bf k}_1,{\bf k}_2) &=& \frac{5}{3} (a_{\rm NL}-1) + {\cal O}(1)   \\
\label{fnlthth}
f_{\rm NL}^{\theta}({\bf k}_1,{\bf k}_2) &=& \frac{5}{3} (a_{\rm NL}-1) + {\cal O}(1)
\ea
are the terms arising from non-Gaussian initial statistics. Here $a_{\rm NL}$ parametrizes  the primordial NG level, according to the notations of 
Ref.~\cite{bartolosig}.   The ${\cal O}(1)$ terms in Eqs.~(\ref{fnldd})-(\ref{fnlthth}) depend on 
the Fourier-space configuration. These terms represent general relativistic horizon-scale 
corrections~\cite{bartolosig} and have proven to be relevant in the description of the clustering of halos in the presence of non-Gaussian initial 
conditions as was recently discussed in \cite{MV2009}.  Nevertheless, for our purposes here, we will deal with the 
primordial NG in the limit $|a_{\rm NL}-1| \gg 1$, so,  these terms become irrelevant and we will keep just the 
constant part of the non-Gaussian terms $f^{\delta, \theta}_{\rm NL} \simeq  5 (a_{\rm NL}-1)/3$. \footnote{Notice also that we are  
neglecting the non-Gaussian contribution arising from the second-order evolution of perturbations during radiation dominance discussed 
in \cite{Bartolo:2006fj} and \cite{Sen}, since the matching at second-order into the radiation era will generate a term which then scale as 
$a(\tau)\propto \tau^2$, thus being subdominant on small scales w.r.t. the terms accounted here.}   
\\
We assume that primordial NG is encoded in the curvature perturbations described through the Bardeen's gauge 
invariant potential  $\Phi$ which, on sub-horizon scales reduces to the peculiar gravitational potential $\phi$ up to a minus sign. 
The second order expansion of the matter density and velocity divergence can be expressed in terms of $\Phi$, obtaining  
\ba
\label{withPhi} 
\delta_{{\bf k}} (\tau)&=& {\cal M}(k, a)\Phi({\bf k})\nonumber\\ 
&+&\int d^3{\bf k}_1d^3{\bf k}_2\,{\cal K}_{\delta}^N({\bf k}_1,{\bf k}_2) {\cal M}(k_1, a)\Phi^{(1)}_{{\bf k}_1} {\cal M}(k_2, a)\Phi^{(1)}_{{\bf k}_2}\delta_D ({\bf k}_{12}-{\bf k})\\ 
\label{Theta} 
 -\frac{\theta_{{\bf k}} (\tau)}{{\cal H} f } &=&  {\cal M}(k, a)\Phi({\bf k}) \nonumber\\ 
&+&  \int d^3{\bf k}_1d^3{\bf k}_2\,{\cal K}_{\theta}^N({\bf k}_1,{\bf k}_2) {\cal M}(k_1, a)\Phi^{(1)}_{{\bf k}_1} {\cal M}(k_2, a)\Phi^{(1)}_{{\bf k}_2}\delta_D ({\bf k}_{12}-{\bf k})\, ,
\ea
where the primordial gravitational potential $\Phi$ includes the linear part and the contributions coming from primordial non-Gaussianities. The function ${\cal M}$ comes from the Poisson equation 
\be
{\cal M}(k, a) = \frac{2 k^2 T(k)}{3\Omega_{m,0}{H}_0^2}D_{+}(a),
\ee
with $T(k)$ being the transfer function of matter fluctuations normalized such that $T(k)\rightarrow 1$ when $k\rightarrow 0$. The expressions above account for a generic shape of the primordial NG, and also for a $\Lambda$CDM universe. Notice that, while for the primordial gravitational potential, the first lines of Eq.~(\ref{withPhi}) and~(\ref{Theta}) are exact, over the second lines applies the same approximations that led to eqs.~(\ref{2ndd})-(\ref{2ndth}). Using the expressions above, and recalling the definition (\ref{phi-dt}) in which we replace $\varphi_2 \rightarrow \varphi_2 /f(\tau)$ in the $\Lambda$CDM case, the initial BS components can be written as

\ba
\left\langle \varphi_1 ({\bf k}_1)\varphi_1 ({\bf k}_2)\varphi_1 ({\bf k}_3)\right\rangle &=&  e^{-3\eta} \delta_D ({\bf k}_1+ {\bf k}_2 +{\bf k}_3) \times \nonumber\\ 
&&\left[ 2 {\cal K}_{\delta}^N({\bf k}_1,{\bf k}_2) {\cal M}^2(k_1,a){\cal M}^2(k_2,a) P_{\Phi}(k_1)P_{\Phi}(k_2)+ cycl. \right] \nonumber\\
\label{bsphi111} 
&+& e^{-3\eta}{\cal M}(k_1,a){\cal M}(k_2,a){\cal M}(k_3,a)\left\langle \Phi_p ({\bf k}_1)\Phi_p ({\bf k}_2)\Phi_p ({\bf k}_3)\right\rangle \\ 
\left\langle \varphi_2 ({\bf k}_1)\varphi_2 ({\bf k}_2)\varphi_2 ({\bf k}_3)\right\rangle &=&  e^{-3\eta} \delta_D ({\bf k}_1+ {\bf k}_2 +{\bf k}_3) \times \nonumber\\ 
&&\left[ 2 {\cal K}_{\theta}^N({\bf k}_1,{\bf k}_2) {\cal M}^2(k_1,a){\cal M}^2(k_2,a) P_{\Phi}(k_1)P_{\Phi}(k_2)+ cycl. \right]\nonumber\\ 
\label{bsphi222}
&+& e^{-3\eta} {\cal M}(k_1,a){\cal M}(k_2,a){\cal M}(k_3,a)\left\langle \Phi_p ({\bf k}_1)\Phi_p ({\bf k}_2)\Phi_p ({\bf k}_3)\right\rangle \\ 
\left\langle \varphi_1 ({\bf k}_1)\varphi_2 ({\bf k}_2)\varphi_2 ({\bf k}_3)\right\rangle &=& 2 e^{-3\eta} \delta_D ({\bf k}_1+ {\bf k}_2 +{\bf k}_3) \times \nonumber\\ 
&&\left[  {\cal K}_{\theta}^N({\bf k}_1,{\bf k}_2) {\cal M}^2(k_1,a){\cal M}^2(k_2,a)P_{\Phi}(k_1)P_{\Phi}(k_2) \right. \nonumber\\ 
&+& \left.{\cal K}_{\theta}^N({\bf k}_1,{\bf k}_3) {\cal M}^2(k_1,a){\cal M}^2(k_3,a)P_{\Phi}(k_1)P_{\Phi}(k_3)\right. \nonumber\\ 
&+&\left. {\cal K}_{\delta}^N({\bf k}_2,{\bf k}_3) {\cal M}^2(k_2,a){\cal M}^2(k_3,a)P_{\Phi}(k_2)P_{\Phi}(k_3) \right]\nonumber\\ 
\label{bsphi122}
&+&e^{-3\eta} {\cal M}(k_1,a){\cal M}(k_2,a){\cal M}(k_3,a)\left\langle \Phi_p ({\bf k}_1)\Phi_p ({\bf k}_2)\Phi_p ({\bf k}_3)\right\rangle \\ 
\left\langle \varphi_1 ({\bf k}_1)\varphi_1 ({\bf k}_2)\varphi_2 ({\bf k}_3)\right\rangle &=& 2 e^{-3\eta} \delta_D ({\bf k}_1+ {\bf k}_2 +{\bf k}_3) \times \nonumber\\ 
&&\left[  {\cal K}_{\theta}^N ({\bf k}_1,{\bf k}_2) {\cal M}^2(k_1,a){\cal M}^2(k_2,a)P_{\Phi}(k_1)P_{\Phi}(k_2) \right. \nonumber\\ 
&+&\left. {\cal K}_{\delta}^N({\bf k}_1,{\bf k}_3) {\cal M}^2(k_1,a){\cal M}^2(k_3,a)P_{\Phi}(k_1)P_{\Phi}(k_3)\right. \nonumber\\ 
&+&\left. {\cal K}_{\delta}^N({\bf k}_2,{\bf k}_3) {\cal M}^2(k_2,a){\cal M}^2(k_3,a)P_{\Phi}(k_2)P_{\Phi}(k_3) \right]\nonumber\\ 
\label{bsphi112}
&+&e^{-3\eta}{\cal M}(k_1,a){\cal M}(k_2,a){\cal M}(k_3,a)\left\langle \Phi_p ({\bf k}_1)\Phi_p ({\bf k}_2)\Phi_p ({\bf k}_3) \right\rangle 
\ea
where $P_\Phi(k)$ is the gravitational potential power-spectrum, see Eq.~(\ref{powerPhi}) below. The expressions above constitute the 
initial conditions for the bispectrum in the system (\ref{system}) and  must be evaluated at the initial time, for which $\eta=\log{a}/{a_{in}}=0$. 
Notice that the expressions above are separated in a perturbative Newtonian term, which is the first term on the r.h.s of each equation, 
and in a primordial part, which is the term including the three-point correlator of the primordial gravitational potential. 

\subsection{Primordial NG: Shapes and Running}
The functional form of the bispectrum of the primordial gravitational potential entering in Eqs.~(\ref{bsphi111})-(\ref{bsphi112}). 
\be
\langle \Phi({\bf k}_1) \Phi({\bf k}_2) \Phi({\bf k}_3) \rangle =  \delta_D ({\bf k}_1+ {\bf k}_2 +{\bf k}_3) B_{\Phi}( k_1,  k_2 ,  k_3)
\ee
depends on the details of the mechanism that generated the primordial fluctuations. In the following we will analyze three 
representative phenomenological models associated to different mechanisms generating the primordial NG. 
Note that there exist other possible configurations (see for instance \cite{Fergusson:2008ra}) which correspond to more general 
deviations from the standard slow-roll inflation and which are not well described by the models treated here.

\subsubsection{Local Shape}
In the local model, the NG for Bardeen's gauge invariant primordial gravitational potential $\Phi$ is generated by a quadratic expansion which is local in real space
\be\label{PhiNG}
\Phi({\bf x}) = \Phi_{\rm L}({\bf x}) + f_{\rm NL}^{\rm local} (\Phi_{\rm L}^2({\bf x})- \langle\Phi_{\rm L}^2\rangle) \;,
\ee
where $\Phi_L$ is the linear Gaussian part of the potential. The dimensionless constant $f_{\rm NL}$ defines the NG strength in this model. 
One must also notice that, since $\Phi$ and $\Phi_L$ evolve proportionally to $g(a)=D_{+}(a)/a$, the Eq.~(\ref{PhiNG}), and consequently, the definition of 
$f_{\rm NL}$, depends on the choice of the redshift at which this equation is extrapolated. The relation between two extrapolation choices $a$ and $b$ 
is expressed through  $f_{\rm NL}^{a} =g(z=z_b)/g(z=z_a) f_{\rm NL}^{b}$. Two common conventions in the literature are the LSS convention in 
which Eq.~(\ref{PhiNG}) is extrapolated to $z=0$, and the CMB convention which  extrapolates to $z=\infty$. They are related through $f_{\rm NL}^{\rm LSS} = 
g(z=\infty)/g(0) f_{\rm NL}^{\rm CMB}$ where $g(z=\infty)/g(0)\simeq 1.3064$. In this paper, we assume the CMB convention, which implies that $\Phi$ is 
evaluated deep in the matter era. From equation (\ref{PhiNG}) one can easily derive the expression for the local bispectrum
\be
B_{\Phi}( k_1,  k_2 ,  k_3) = f_{\rm
  NL}^{\rm local}  F^{\rm local}( k_1,  k_2 ,  k_3),
\ee
where 
\ba
\label{eq:local}
 F^{\rm local}( k_1,  k_2 ,  k_3) &=&  2 \left[P_\Phi(k_1) P_\Phi(k_2) +P_\Phi(k_1) P_\Phi(k_3) + P_\Phi(k_2) P_\Phi(k_3)
\right] \\  &=&  2 \Delta_\Phi^2
\cdot \left(\frac1{(k_1 k_2)^{3-(n_s-1)}} + \frac1{(k_1 k_3)^{3-(n_s-1)}} + 
\frac1{(k_2 k_3)^{3-(n_s-1)}}\right)\nonumber
\ea
and $P_{\Phi}(k)$ is the gravitational potential power spectrum 
\be
\label{powerPhi}
\langle \Phi({\bf k}_1) \Phi({\bf k}_2)
\rangle \equiv  \delta^3\big({\bf k}_{12}\big) P_\Phi(k_1) =  \delta^3\big({\bf k}_{12}\big)\Delta_\Phi \cdot k^{-3+(n_s-1)}
\ee
with scalar amplitude $\Delta_\Phi$  and tilt $n_s$. In slow-roll single scalar  field inflation, $f_{\rm NL}^{\rm local} $ is suppressed by the 
slow-roll parameters and consequently the primordial NG is unmeasurably small \cite{acquaviva,maldacena}. 
However, in the presence of light fields other than the inflaton, as for instance in the curvaton model or in multi-field models, inflation can produce 
large NG of the local type with $| f_{\rm NL}^{\rm local} | \sim {\cal O}(10^2)$~\cite{bartoloreview}. 
The strongest current
limits on the strength of local NG set the $f^{\rm loc}_{\rm NL}$
parameter to be in the range $-4<f^{\rm loc} _{\rm NL}<80$ at 95\% confidence level
\cite{zal}. 
\subsubsection{Equilateral Shape}
We can also consider inflationary models with higher order derivative operators and non canonical kinetic terms, such as the DBI model~\cite{DBI}, for instance. In this case, as was discussed in \cite{Babich:2004gb},
the primordial bispectrum is well described by the {\it equilateral} template $B_{\Phi}( k_1,  k_2 ,  k_3) = f_{\rm  NL}^{\rm eq}  F^{\rm eq}( k_1,  k_2 ,  k_3)$, where  
\ba
\label{eq:ours}
\nonumber F^{\rm eq}(k_1,k_2,k_3) & = & 6 \left(-P_\Phi(k_1)P_\Phi(k_2) +
   (2 \; {\rm perm}.) - 2\left\{P_\Phi(k_1)P_\Phi(k_2)P_\Phi(k_3)\right\}^{2/3} \right. \\ 
   \nonumber & & \left. +
   \left(P_\Phi(k_1)\left\{P_\Phi(k_2)\right\}^2 \left\{P_\Phi(k_3)\right\}^3\right)^{1/3}
+ (5 \; {\rm perm}.) \right)\\ 
\nonumber & = &   
6 \Delta_\Phi^2 \cdot \left(-\frac1{(k_1 k_2)^{3-(n_s-1)}} -
   (2 \; {\rm perm}.) - \frac2{(k_1 k_2 k_3)^{2-\frac23(n_s-1)} } \right. \\  & & \left. +
   \frac1{k_1^{1-\frac13(n_s-1)} k_2^{2-\frac23(n_s-1)}  k_3^{3-(n_s-1)}}
+ (5 \; {\rm perm}.) \right) \;.
\ea
The primordial bispectrum is maximized by configurations with modes of similar momentum scales $ k_1 \approx k_2 \approx k_3$, i.e. for equilateral configurations.  At present 
$f_{\rm NL}^{\rm eq}$ is constrained to be $-125 < f_{\rm NL}^{\rm eq} < 435 $ ($95 \%$ C.L.) \cite{zal}

\subsubsection{Folded Shape}
\begin{figure}[t] 
\includegraphics[width=8.0in]{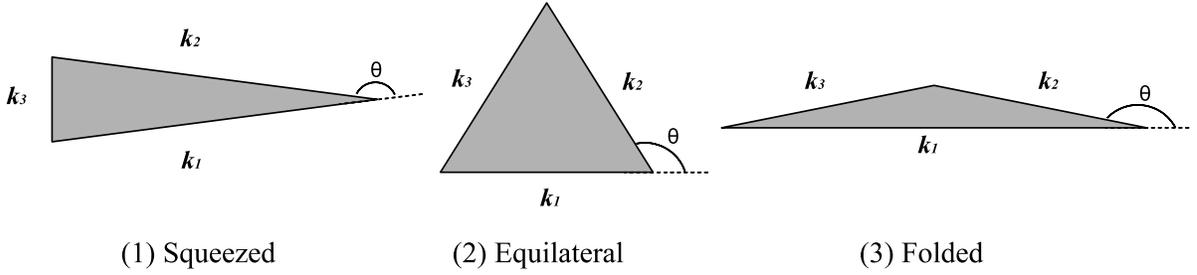} 
	    \caption{Momentum triangle configurations which maximize the templates functions $F(k_1 , k_2 , k_3)$. (1) The squeezed configuration 
	    $k_3 \ll k_1 \approx k_2$ maximize the template function $F^{\rm local}(k_1 , k_2 , k_3)$ produced in local models of inflation. (2) The 
	    equilateral configuration $ k_1 \approx k_2 \approx k_3$ maximize $F^{\rm eq}(k_1 , k_2 , k_3)$ which arises on models with higher-order 
	    derivatives and non canonical terms. (3) The ``folded" or ``flattened" configuration $ k_2 \approx k_3 \approx  k_{1} /2$ maximize the 
	    $F^{\rm fol.}(k_1 , k_2 , k_3)$ which is related to the choice of a non-standard vacuum other than the Bunch-Davies vacuum. }     \label{shapes}
\end{figure}

A third template is the so called {\it folded} or {\it flattened} model. This model is related to the non-Gaussianities generated by  deviations coming 
from the choice of a non-standard adiabatic Bunch-Davies vacuum state as initial state \cite{Chen:2006nt,HolmanTolley}. The associated primordial bispectrum 
assumes a complicated functional form in terms of the momenta $k_{i}$. Nevertheless, in Ref.~\cite{Meerburg:2009ys} it is proposed a factorized 
template which reflects very well the main features of the primordial NG associated to the choice of a non standard vacuum. The proposed 
factorized template is $B_{\Phi}( k_1,  k_2 ,  k_3) = f_{\rm  NL}^{\rm fol.}  F^{\rm fol.}( k_1,  k_2 ,  k_3)$, with
\ba
\label{fld} 
\nonumber F^{\rm fol.}(k_1,k_2,k_3) & = & 6\left(P_\Phi(k_1)P_\Phi(k_2) +
   (2 \; {\rm perm}.) + 3\left\{P_\Phi(k_1)P_\Phi(k_2)P_\Phi(k_3)\right\}^{2/3} \right. \\ 
   \nonumber & & \left. -
   \left(P_\Phi(k_1)\left\{P_\Phi(k_2)\right\}^2 \left\{P_\Phi(k_3)\right\}^3\right)^{1/3}
+ (5 \; {\rm perm}.) \right)\\ 
\nonumber & = &    
6 \Delta_\Phi^2 \cdot \left(\frac1{(k_1 k_2)^{3-(n_s-1)}} +
   (2 \; {\rm perm}.) + \frac3{(k_1 k_2 k_3)^{2-\frac23(n_s-1)} } \right. \\  & & \left. -
   \frac1{k_1^{1-\frac13(n_s-1)} k_2^{2-\frac23(n_s-1)}  k_3^{3-(n_s-1)}}
- (5 \; {\rm perm}.) \right) \;.
\ea
The primordial bispectrum for this model is maximized by configurations with modes obeying the momentum configuration $ k_2 \approx k_3 \approx  k_1 /2$. 
The triangle configurations relevant for each model of NG are represented in figure \ref{shapes}.

\subsubsection{Running $f_{\rm NL}$}
So far, we have considered a scale independent $f_{\rm NL}$ parameter regardless the shape of the primordial bispectrum. However, recently, several inflationary scenarios have been considered which naturally allow for  NG with a generic scale dependence on $f_{\rm NL}$. More specifically, it was shown in Refs.~\cite{Chen:2006nt} and~\cite{Chen:2005fe} that single scalar field models of inflation with variable {\it speed of sound} $c_s$, predict in general a 
primordial bispectrum of the equilateral type with a scale dependent $f_{\rm NL}$.  In the following, for non-Gaussianities of the equilateral type, we adopt the functional form for the scale dependence proposed in \cite{LoVerde:2007ri}:
\be\label{fk}
f_{\rm NL}(k_1, k_2, k_3) = f_{{\rm NL}, P} \left(\frac{k_1 + k_2 + k_3}{3 k_{P}}\right)^{-2\kappa},
\ee
where $f_{{\rm NL}, P}$ is the non linear parameter evaluated at some pivot scale $k_{P}$ and $\kappa$ is a free parameter (related to the speed of sound) 
that quantifies the running on the scale and which is constant at least between CMB and cluster scales \cite{LoVerde:2007ri}. To be consistent with observational 
constraints, the scale dependence must be very mild, then $ |\kappa| \ll 1 $. Also, in an interesting discussion regarding the scale dependence of the biasing 
parameters for the equilateral model, in Ref. \cite{Taruya:2008pg}, the authors found the constraint  $|\kappa| \lesssim 0.3$ in order to guarantee the 
convergence of the integrals quantifying the scale dependence of the biasing. In this paper we thus consider values of the 
running parameter within $|\kappa| \leq 0.3$. Moreover, we use negatives values of $\kappa$ since in this case the NG is enhanced for 
scales beyond the pivot scale~\cite{LoVerde:2007ri}. Another proposal for the functional dependence on the running $f_{\rm NL}$ was discussed recently in \cite{Sefusatti:2009xu}
\be\label{fk2}
\bar{f}_{\rm NL}(k_1, k_2, k_3) = f_{{\rm NL}, P} \left(\frac{(k_1  k_2  k_3)^{1/3}}{k_{P}}\right)^{-2\kappa},
\ee
which replaces the {\it arithmetic mean} of the momenta in equation (\ref{fk}) with their {\it geometric mean}. Notice that the two functions coincide for equilateral configurations. 
While the last proposal has the virtue of being separable, which simplifies significantly the numerical implementation of the CMB estimator, 
the equation (\ref{fk}) provides a more accurate description of the primordial bispectrum in DBI models. The functional dependence (\ref{fk2}) 
seems to be better suited for NG of the local type.  It was shown recently that 
non-Gaussianities of the local type exhibit significant scale dependence of the amplitude of the primordial bispectrum~\cite{Byrnes:2008zy,Byrnes:2009pe}. The form of this 
amplitude depends on the specific details of the underlying inflationary model. For example in the case of standard single-field models of inflation it 
coincides with the functional form (\ref{fk2}) in the equilateral limit. Away from this limit, it is not possible to factor out the 
geometric mean of the momenta in $f_{\rm NL}$ but, in principle, the essentials of the local model can be roughly captured with the expression~(\ref{fk2}).

\section{Results}
For our analysis, we assume cosmological parameters corresponding to a $\Lambda$-CDM model in agreement with the 
WMAP 5-year  data \cite{Komatsu:2008hk}. The cosmological parameters for this model are summarized in Table 1. 
As discussed before, we use second order perturbation theory from $z=\infty$ to $z=z_{in}$, and the TRG from $z_{in}$ to $z=0$. 
To get the numerical results presented in the following, we set $z_{in}=50$, and used the CAMB code \cite{Lewis:1999bs} to derive the linear power-spectrum.
In order to reproduce the cosmological model in table 1, we must fix the scalar amplitude to be $\Delta^2_{\cal R}=2.48\times 10^{-9}$ in the notations of Ref.~\cite{Komatsu:2008hk}.
With these parameters we are able to compare our results for the matter power spectrum with the results of N-body simulation of cosmological 
structure formation with non-Gaussian initial conditions of the local type from Refs.~\cite{Pillepich:2008ka} and~\cite{Giannantonio:2009ak} (see also \cite{grossi,Dalal,des,ko} for 
N-body simulations with local-type non-Gaussian initial conditions). 

We also compute the reduced bispectrum for matter density perturbations. 
Here we point out some general features about the effect of the primordial non-Gaussianities on the bispectrum based on our results. 
\begin{table}[h]\label{lcdm5}
  \caption{$\Lambda$-CDM cosmological parameters assumed on this paper.}
\begin{center} 
\begin{tabular}{ | c | c | c | c | c | c | c |}
    \hline
    Model & $h$ & $\sigma_8$  & $n_s$  & $\Omega_m$ & $\Omega_{\Lambda}$ & $\Omega_{b}$ \\ \hline
    $\Lambda$-CDM WMAP5 & 0.701 & 0.817 & 0.96 & 0.279 & 0.721 & 0.0462\\
    \hline    
    \end{tabular}
    \end{center}
\end{table}\\
\subsection{The Power Spectrum}
Here, we present the results of the power spectrum of the matter density perturbations resulting from the integration of the system (\ref{system}) 
on the scales of interest for next generation galaxy surveys. In order to evaluate the impact of primordial non-Gaussianities on the power-spectrum 
we plot the ratio of the power spectra for non-Gaussian and Gaussian initial conditions for different values of the $f_{\rm NL}$ parameter. 
It must be noticed that the power spectra for both Gaussian and non-Gaussian initial conditions were calculated using the TRG method.

\subsubsection{Local Model}
\begin{figure}[h] 
\includegraphics[width=3.1in]{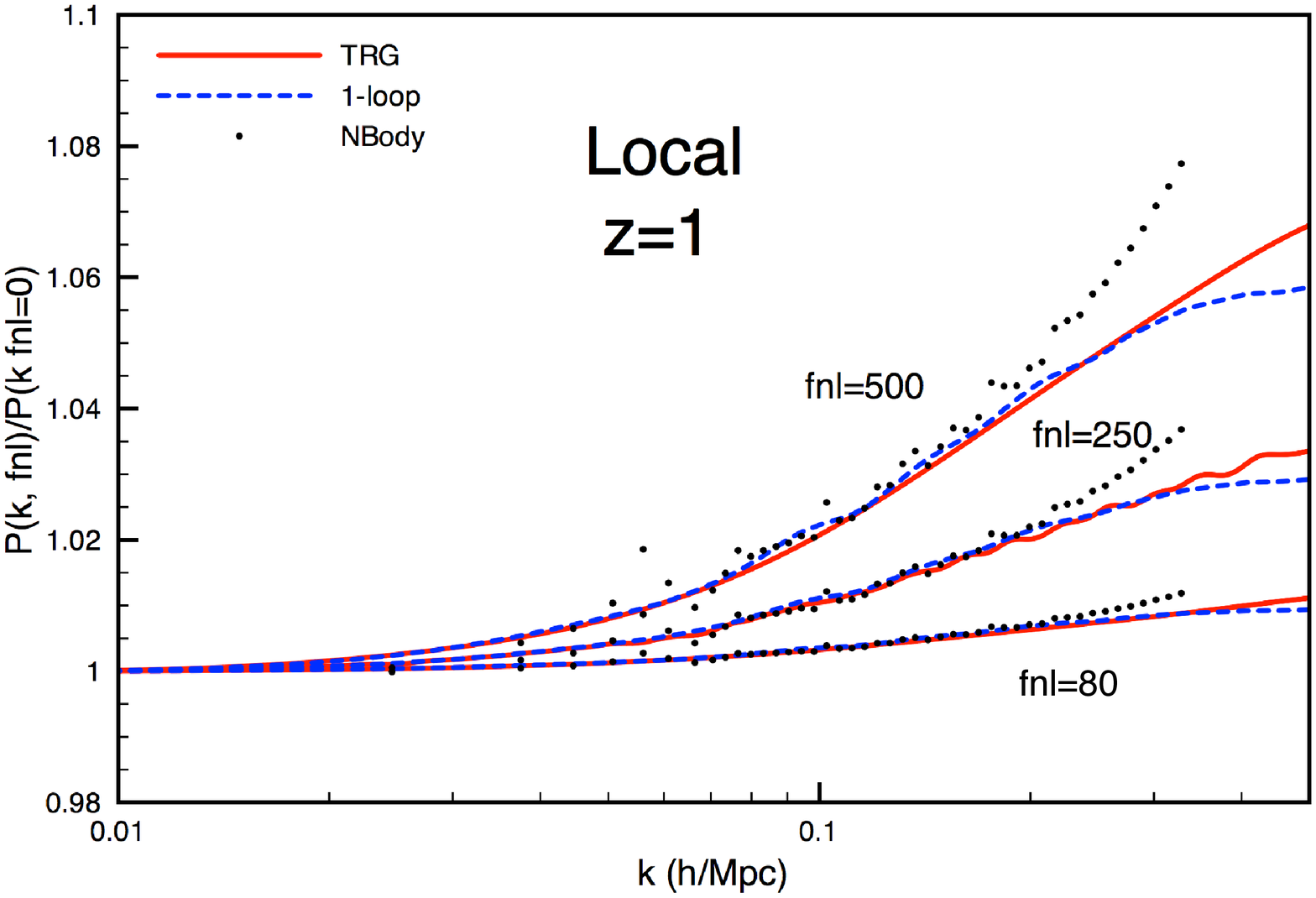} \includegraphics[width=3.1in]{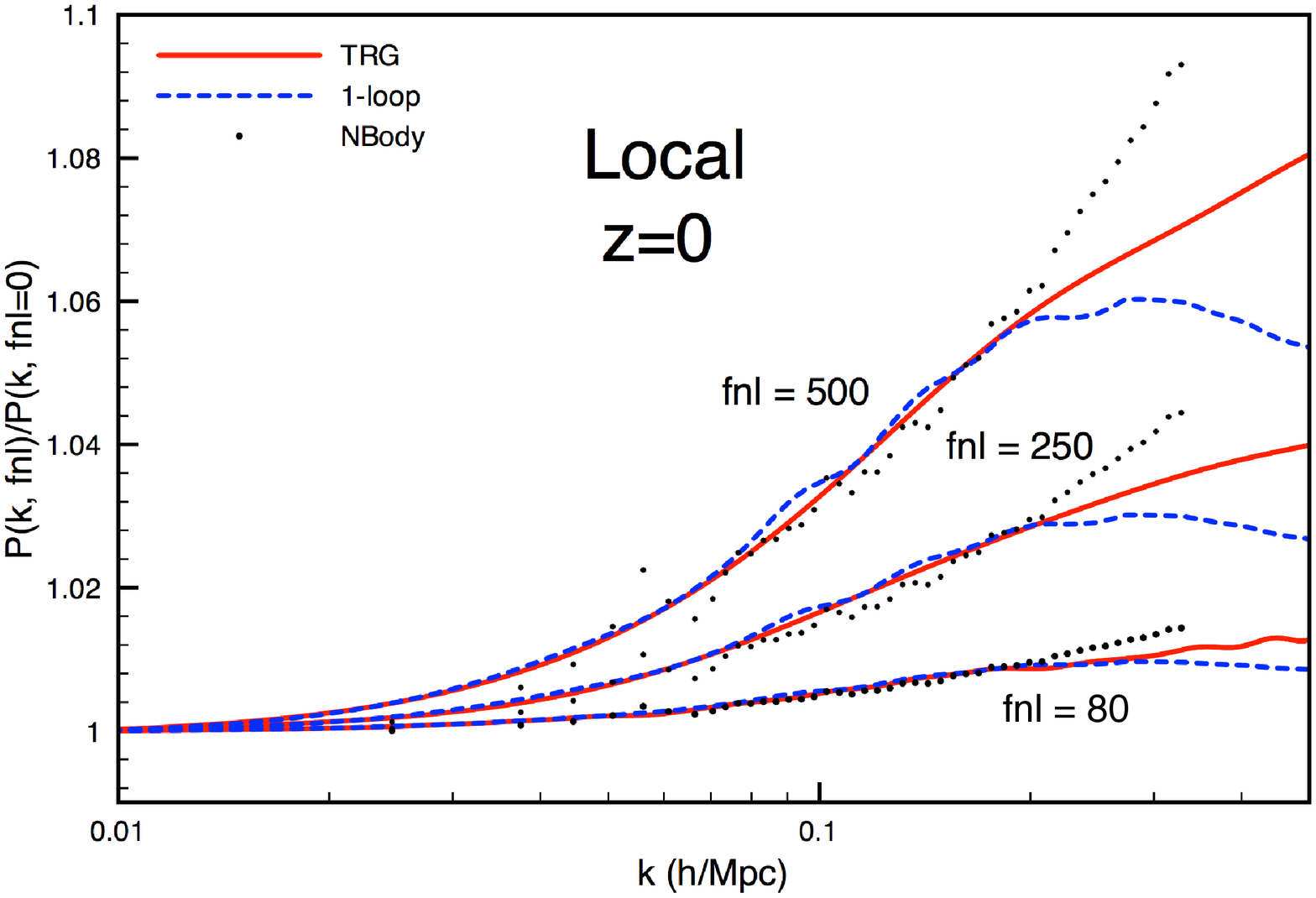} 
    \caption{Ratio of the non-Gaussian to Gaussian power spectrum for several values of $f_{\rm NL}$ in the local model. The dots correspond to the data from the 
    N-body simulations of \cite{Pillepich:2008ka}. The red (continuous) line is the TRG result of this paper and the blue (dashed) line is the one-loop result. }      \label{ratiolocal}
\end{figure}

For the local non-Gaussian model, the result of this evaluation is shown in figure \ref{ratiolocal}. For this type of NG we can compare our findings
with the results of N-body simulations in \cite{Pillepich:2008ka}. In the figures we also show the result from one-loop perturbation theory, see for instance 
 \cite{Taruya:2008pg}. We notice that up to $k\sim 0.2\,h$/Mpc, the one-loop and the TRG describes the data very  accurately while, for larger values of $k$  
we begin to have departures from the data. The difference between one-loop and TRG prediction becomes noticeable for large $f_{\rm NL}$ and for 
higher wavenumbers and lower redshifts as evidenced in the right panel of Figure \ref{ratiolocal}, which is the result of the evaluation at $z=0$. 
From the figures we can also infer that the TRG approach allows a less suppressed non-linear growth of the power spectrum in the region $k \gtrsim 0.25\, h$/Mpc. 
This tendency seems to be general for the different models of primordial NG. The oscillatory behaviour exhibited in the TRG plots are the result of numerical errors in the numerical integration of the system \ref{system}. 

\subsubsection{Equilateral and Folded Shapes}
\begin{figure}[h] 
\includegraphics[width=3.1in]{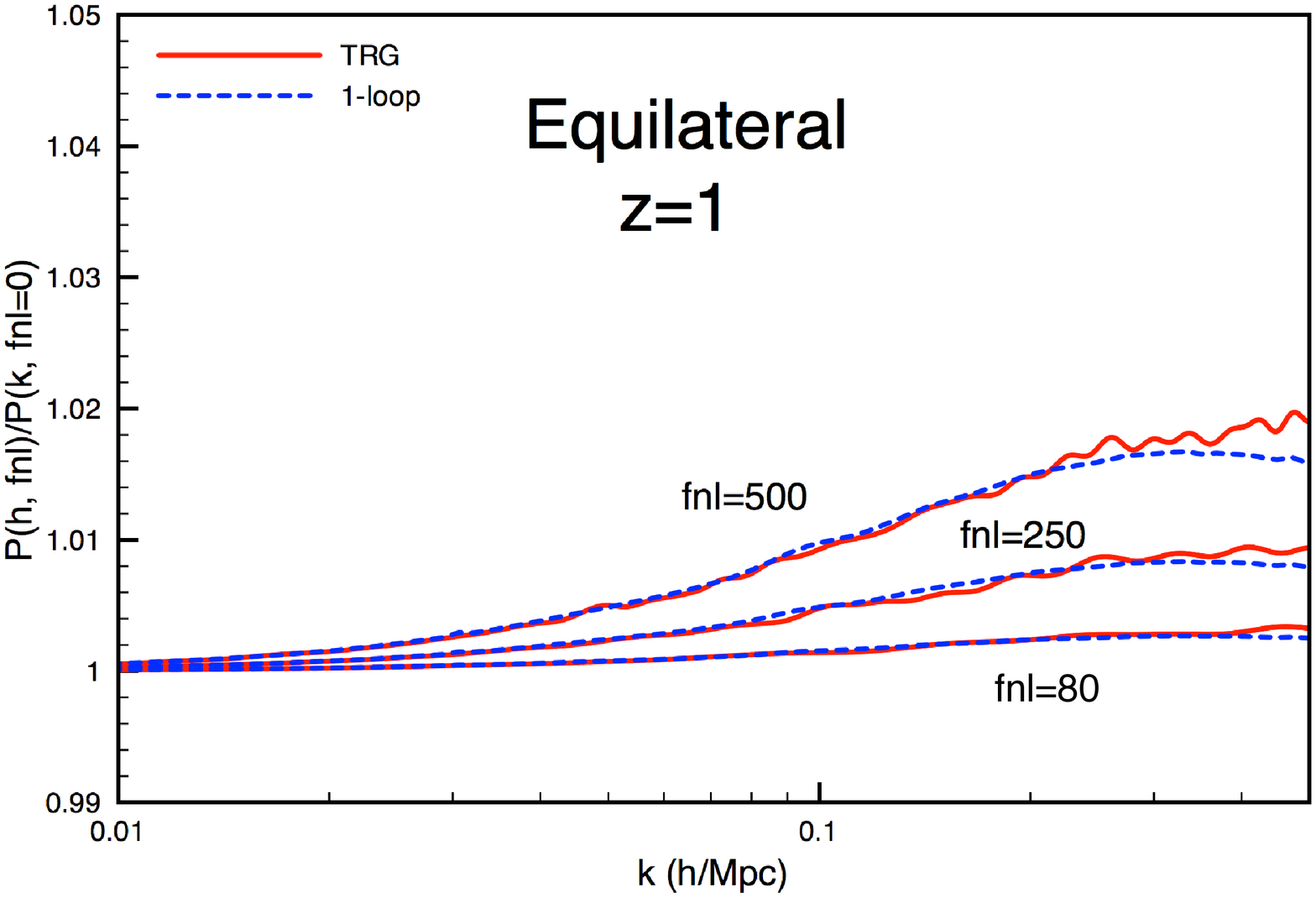} \includegraphics[width=3.1in]{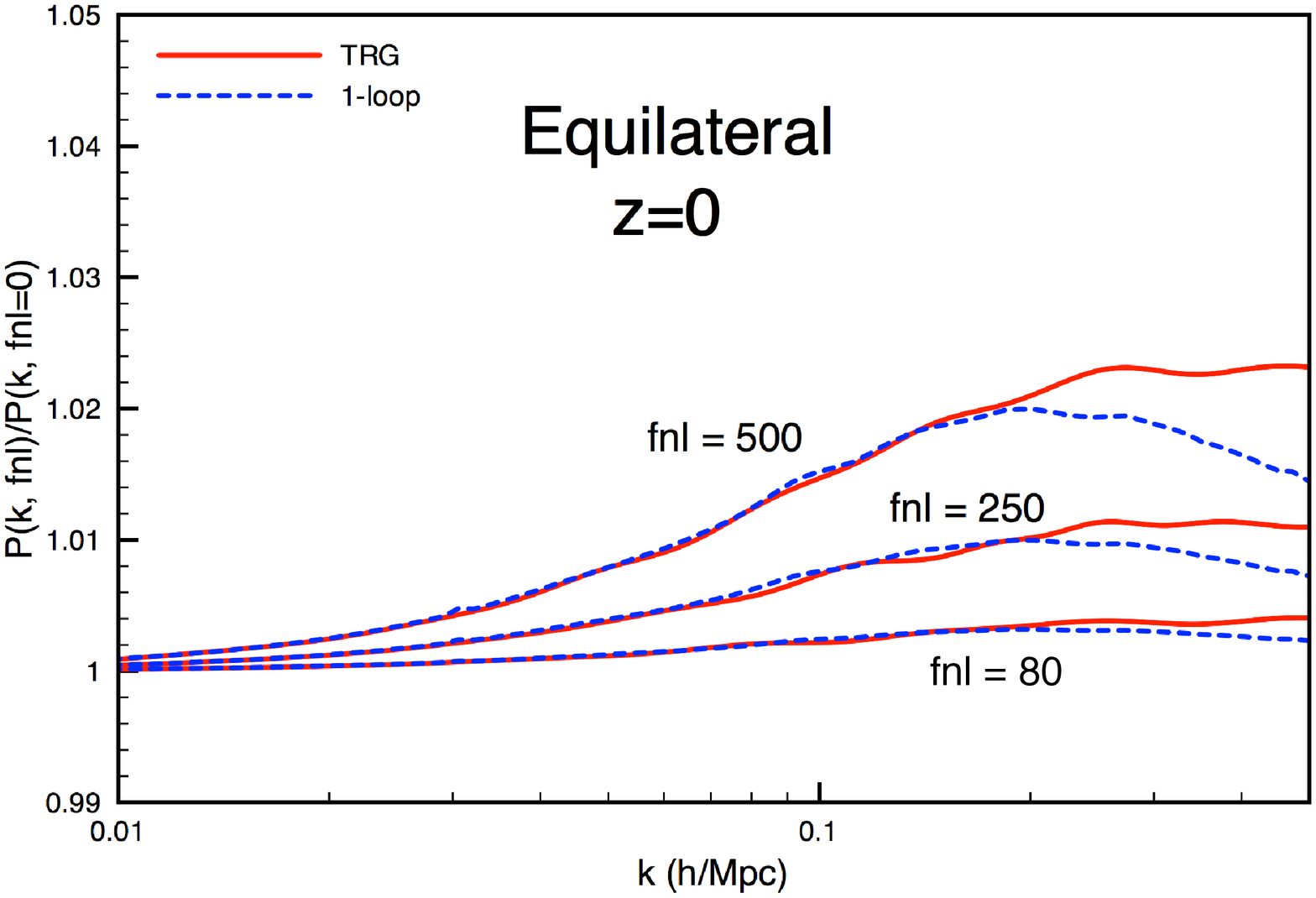} \\
\includegraphics[width=3.1in]{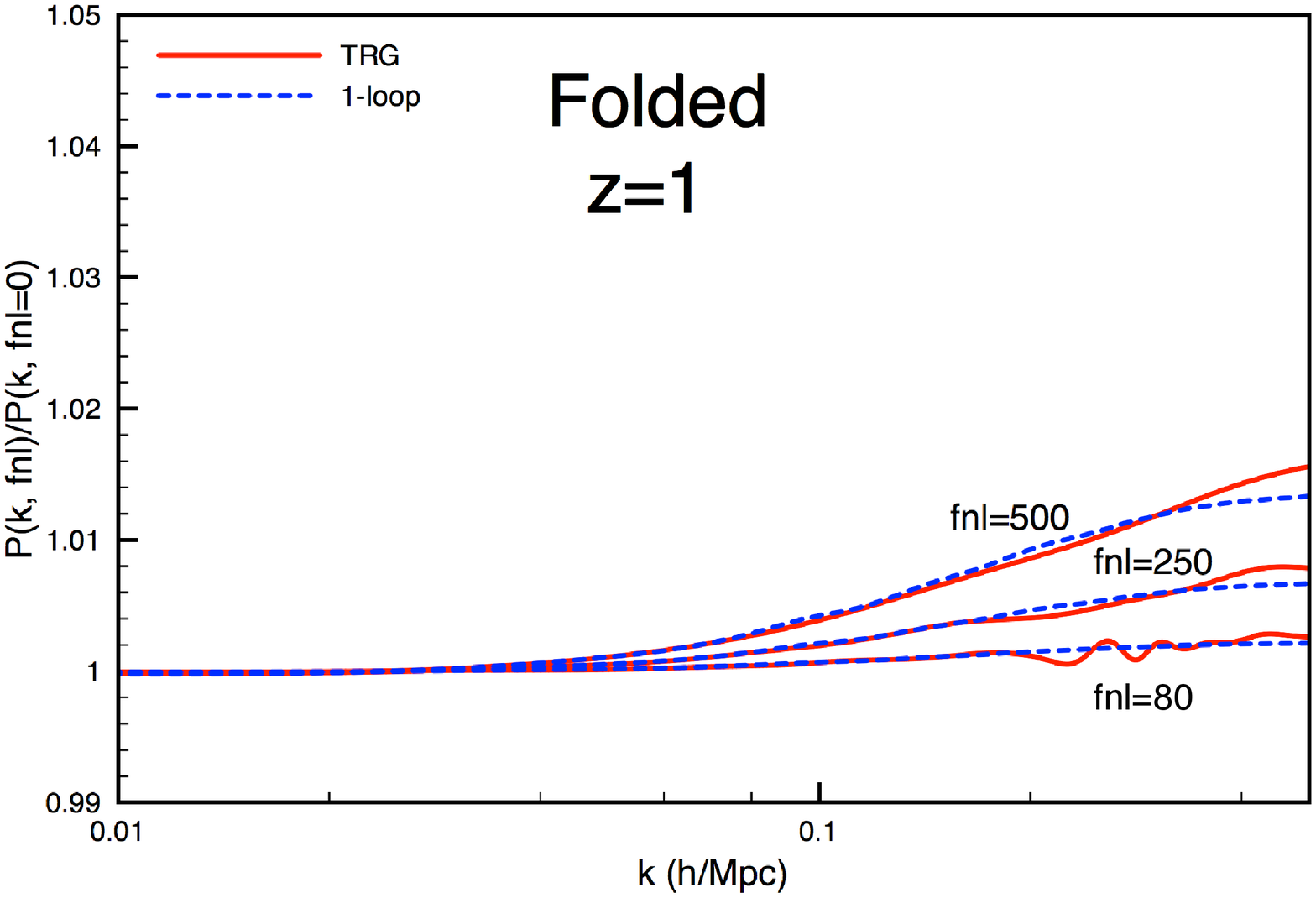} \includegraphics[width=3.1in]{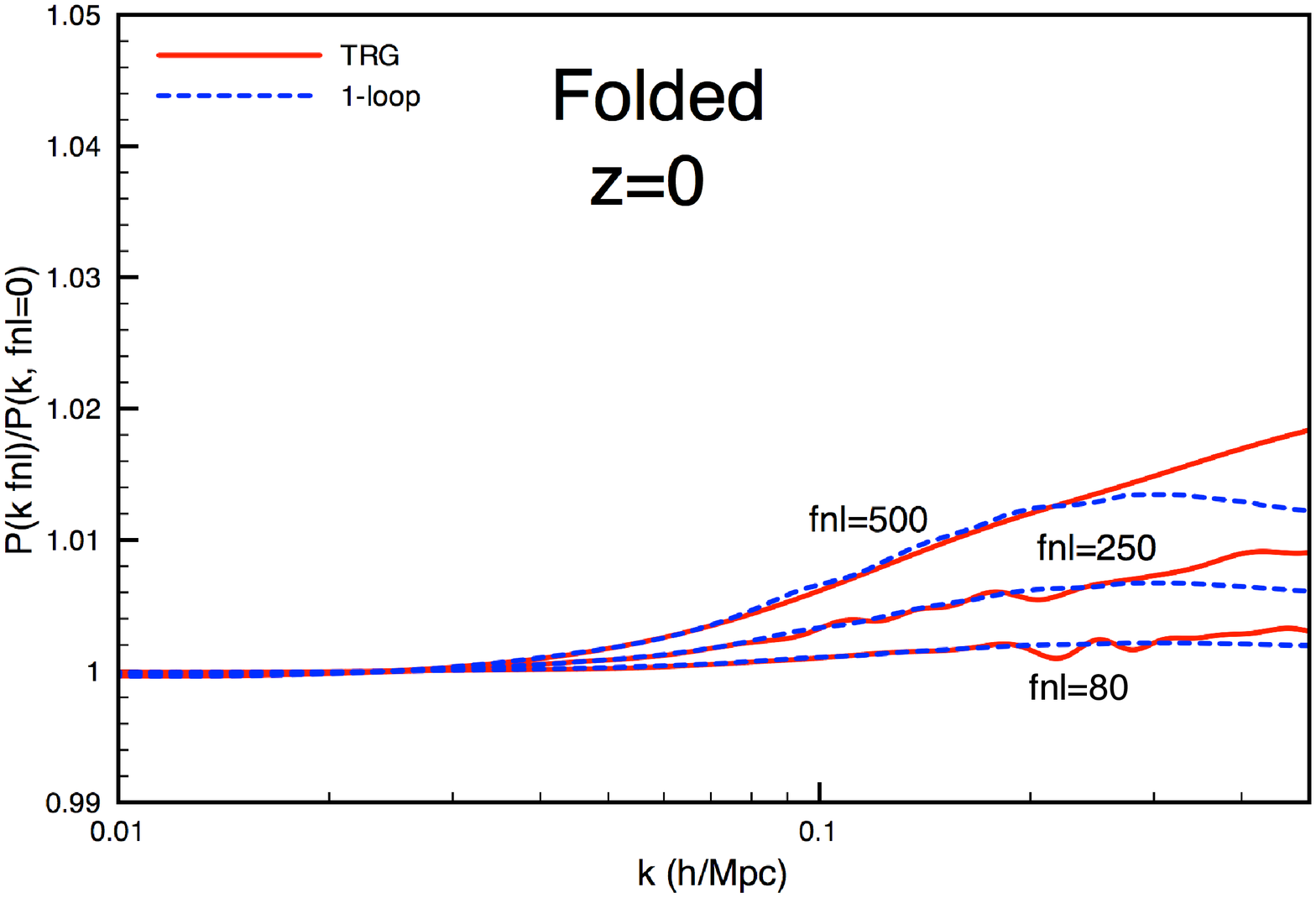}
    \caption{Ratio of the non-Gaussian to Gaussian power spectrum for several values of $f_{\rm NL}$ in the equilateral (top panels) and folded (bottom panels) models. 
    The red (continuous) lines are the TRG result of this paper and the blue (dashed) lines are the one-loop result.}      \label{ratioequifol}
\end{figure}

Here, we show the results of the power spectra ratio for equilateral and folded non-Gaussian shapes. The result of this evaluation is shown in Figures \ref{ratioequifol}. 
Since at present there are no simulations using this type of primordial NG, the results displayed here have a predictive character. 
Again, as we observed in the local case, when we compare with the one-loop calculation the TRG predictions  allow for less suppressed 
effects of non-Gaussianities, The difference is more evident for scales smaller than $k\sim 0.2\,h$/Mpc and for lower redshifts. 

\subsubsection{Running $f_{\rm NL}$, equilateral model}
\begin{figure}[h] 
   \centering
\includegraphics[width=3.1in]{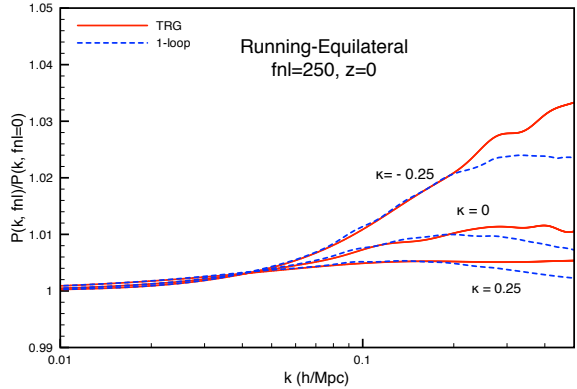}
    \caption{Ratio of the non-Gaussian to Gaussian power spectrum for  $f_{\rm NL}=250$ in the equilateral model for different values of $\kappa$. 
    The red (continuous) lines are the TRG result of this paper and the blue (dashed) lines are the one-loop result.}      \label{ratioRequi}
\end{figure}

In Figure \ref{ratioRequi} we plot the power spectrum ratio for the equilateral model with scale dependent  $f^{\rm eq}_{\rm NL}$ according to Eq.~(\ref{fk}). We choose the pivot scale $k_P= 0.04\, h$/Mpc, 
and $\kappa = 0, 0.25, -0.25$. We see that for positive $\kappa$ we have a suppression in the power spectrum, while, for negative $\kappa$ we can get a significant enhancement of the power spectrum.  
Notice that the scale dependence of $f^{\rm eq}_{\rm NL}$ can make the effects of primordial NG in the power spectrum of the same order of the local case.

\section{The Reduced Bispectrum}
As we will be mainly interested in the study of the dependence of the bispectrum on the shape of the momentum triangle, it is convenient to focus the analysis 
on the {\it reduced bispectrum} which is a useful quantity defined as\footnote{For brevity, in the following we omit the time dependence and the subscripts 
in the correlators since we will deal only with the matter density component.}
\be\label{Q}
Q(k_1, k_2, k_3) \equiv \frac{B(k_1, k_2, k_3)}{\Sigma(k_1, k_2, k_3)},
\ee
where $\Sigma(k_1, k_2, k_3)= P(k_1)P(k_2)+P(k_1)P(k_3)+P(k_2)P(k_3)$. The greatest challenge in the interpretation of galaxy clustering data from any surveys is galaxy bias. 
As thoroughlly
discussed in Ref. \cite{ks},  the reduced  bispectrum provides an excellent determination of linear and non-linear bias parameters of intermediate and high-redhift galaxies, when all measurable triangle configurations down to mildly non-linear scales are included. The reduced bispectrum is also a powerful probe of primordial NG. The planned galaxy surveys at redshift $z>2$ should yield constraints on NG that are comparable to, or even better than, those from CMB experiments.
The precise computation of the reduced bispectrum is therefore of primary importance.

Due to statistical homogeneity and isotropy, the bispectrum and the reduced bispectrum depends on time, on the magnitude of two of the momenta, for 
instance, $k_1$ and $k_2$, and the angle between them $\cos\theta = \hat{k}_1 \cdot \hat{k}_2$. Figure \ref{shapes} shows the geometric parameters 
in the three configurations that we consider in this paper. 
For gaussian initial conditions and an equilateral configuration,  $Q(k,k,k)\simeq 0.57$ at the tree-level in perturbation theory and is
independent of scales. On the other hand, $Q(k,k,k)$ exibits a clear scale dependence when some 
NG is present. The same is true when one departs from the equilateral configuration which reflects the anisotropy on the growth of structures dictated by the non linear Eulerian dynamics.
As the NG appears in our RG equations as post-Newtonian term and therefore suppressed 
at small scales and late times compared to the leading Newtonian terms, it is expected that the NG will be more easily testable at high redshifts and on large 
scales. This is particularly true for the local shape as the reduced bispectrum (at the tree-level) which
is directly proportional to $f_{\rm NL}^{\rm loc}$ scales like $1/{\cal M}(k,a)$. 

In our analysis we fix the ratio $k_{2}/k_{1}=2$ and 
evaluate for different values of $k_{1}$, $f_{\rm NL}$ and for different redshifts. As an extreme case, we also consider the ratio $k_{2}/k_{1}=1$ to have some indications of the maximization of the non-Gaussian effects in the squeezed limit. 

\subsection{Local model}
\begin{figure}[h] 
\includegraphics[width=3.in]{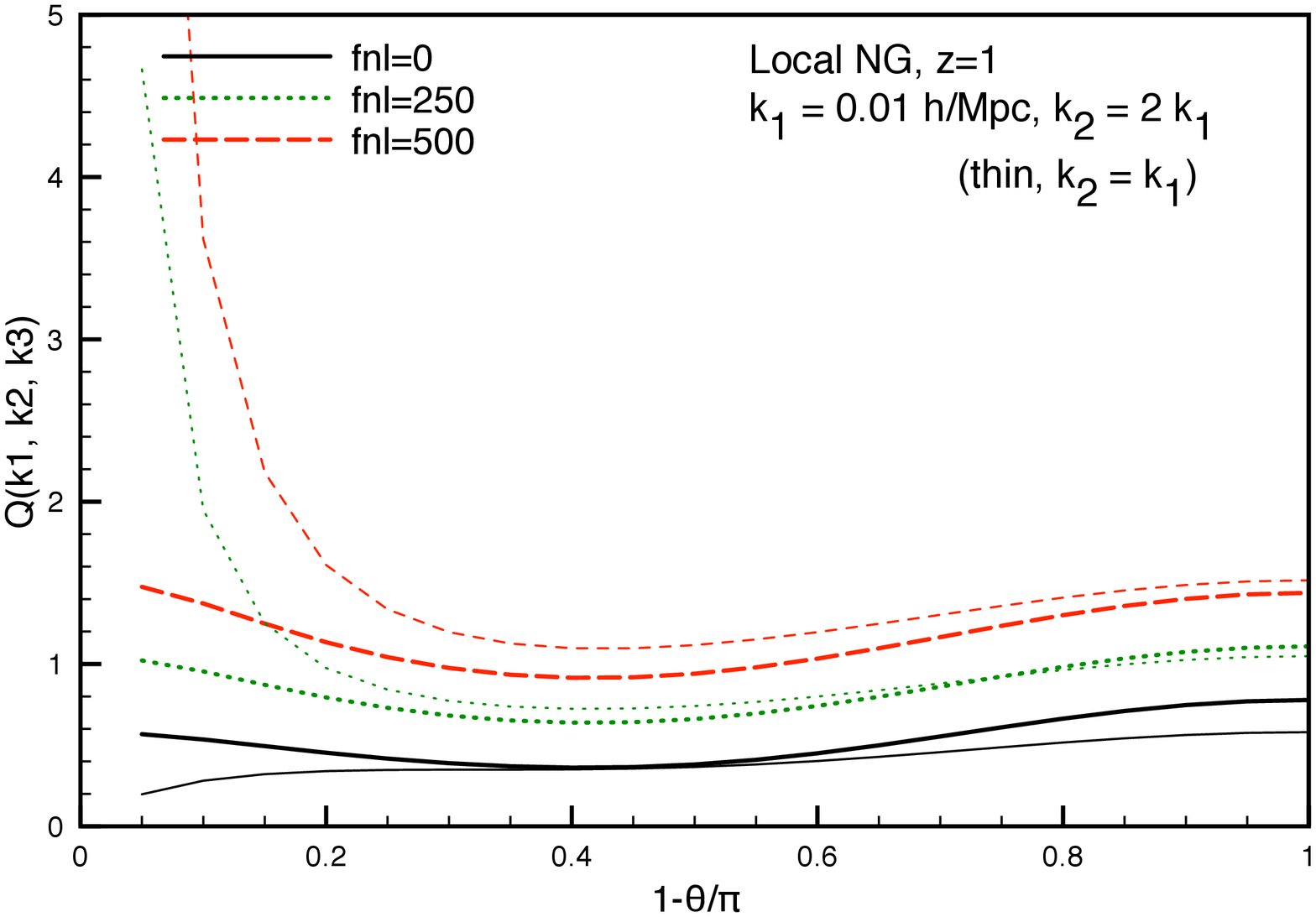} \includegraphics[width=3.in]{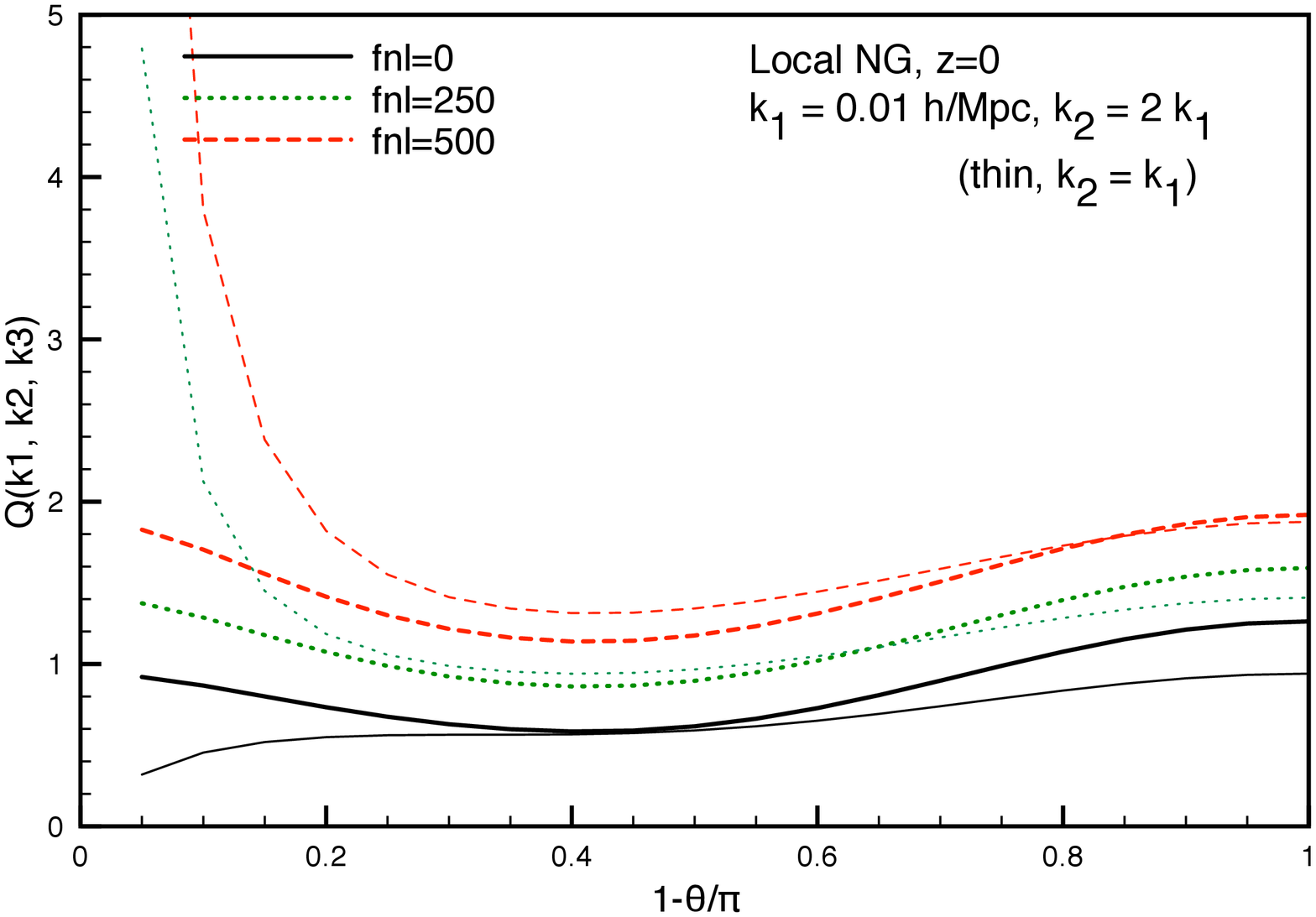} \\
\includegraphics[width=3.in]{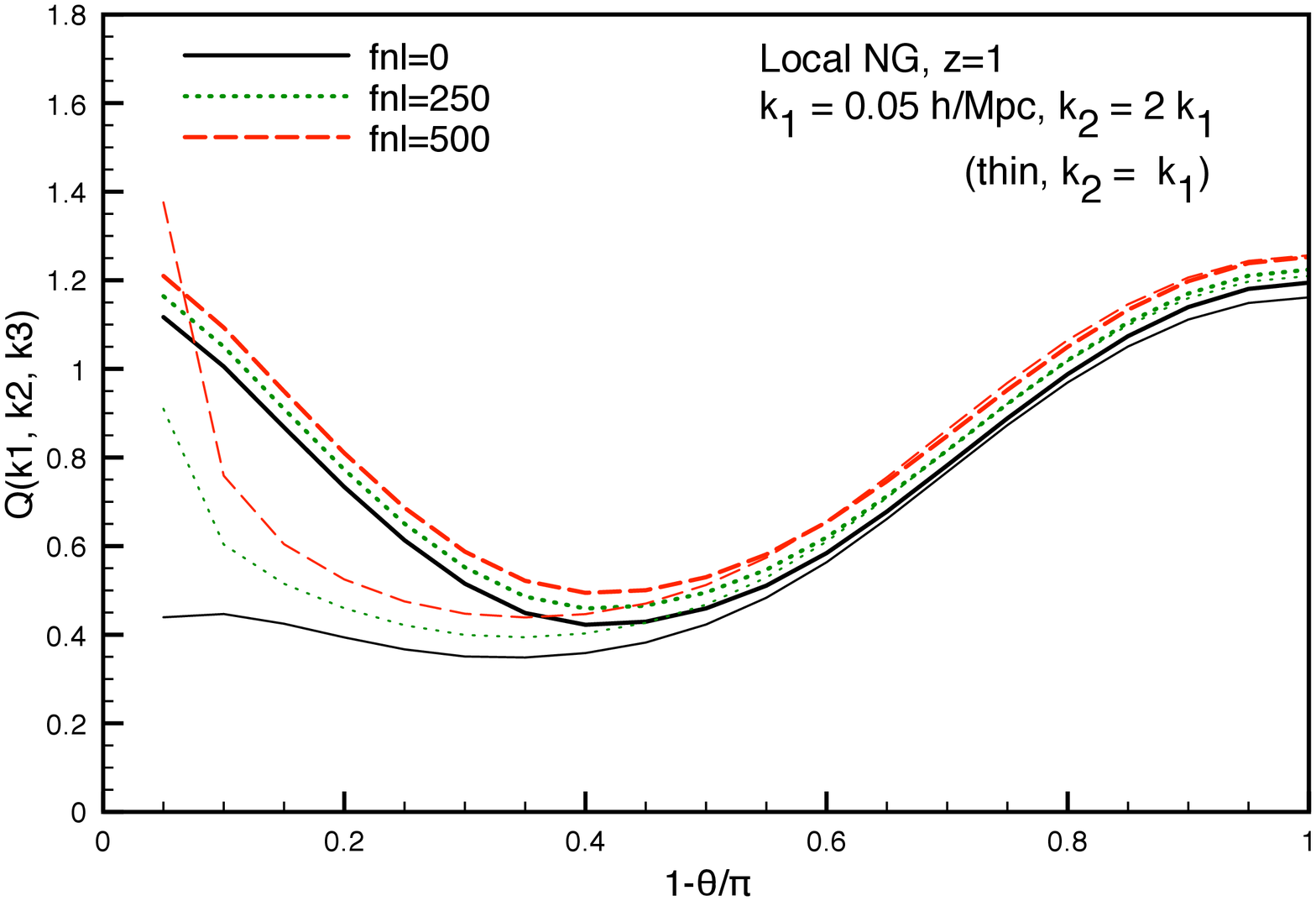} \includegraphics[width=3.in]{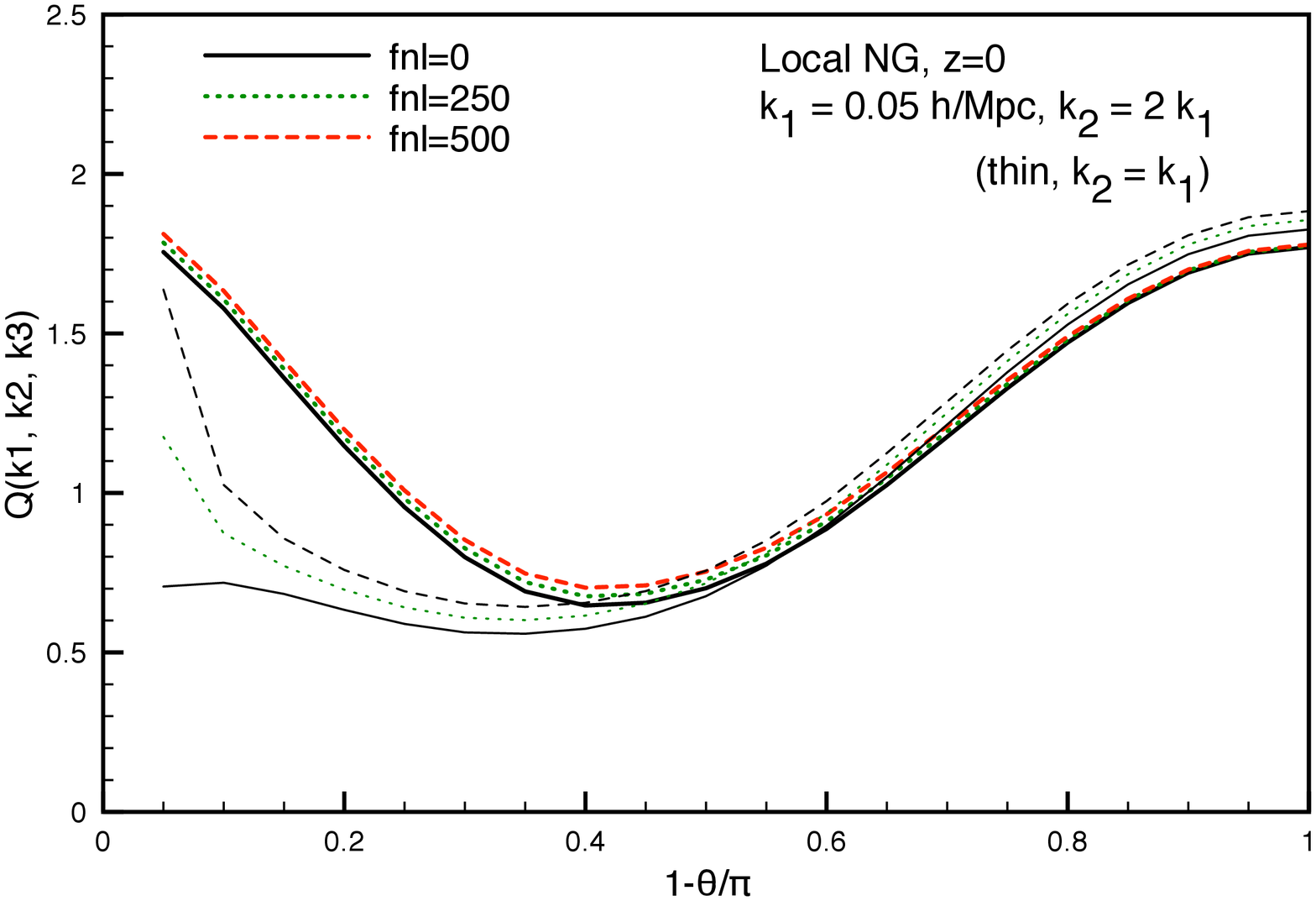} \\
\includegraphics[width=3.in]{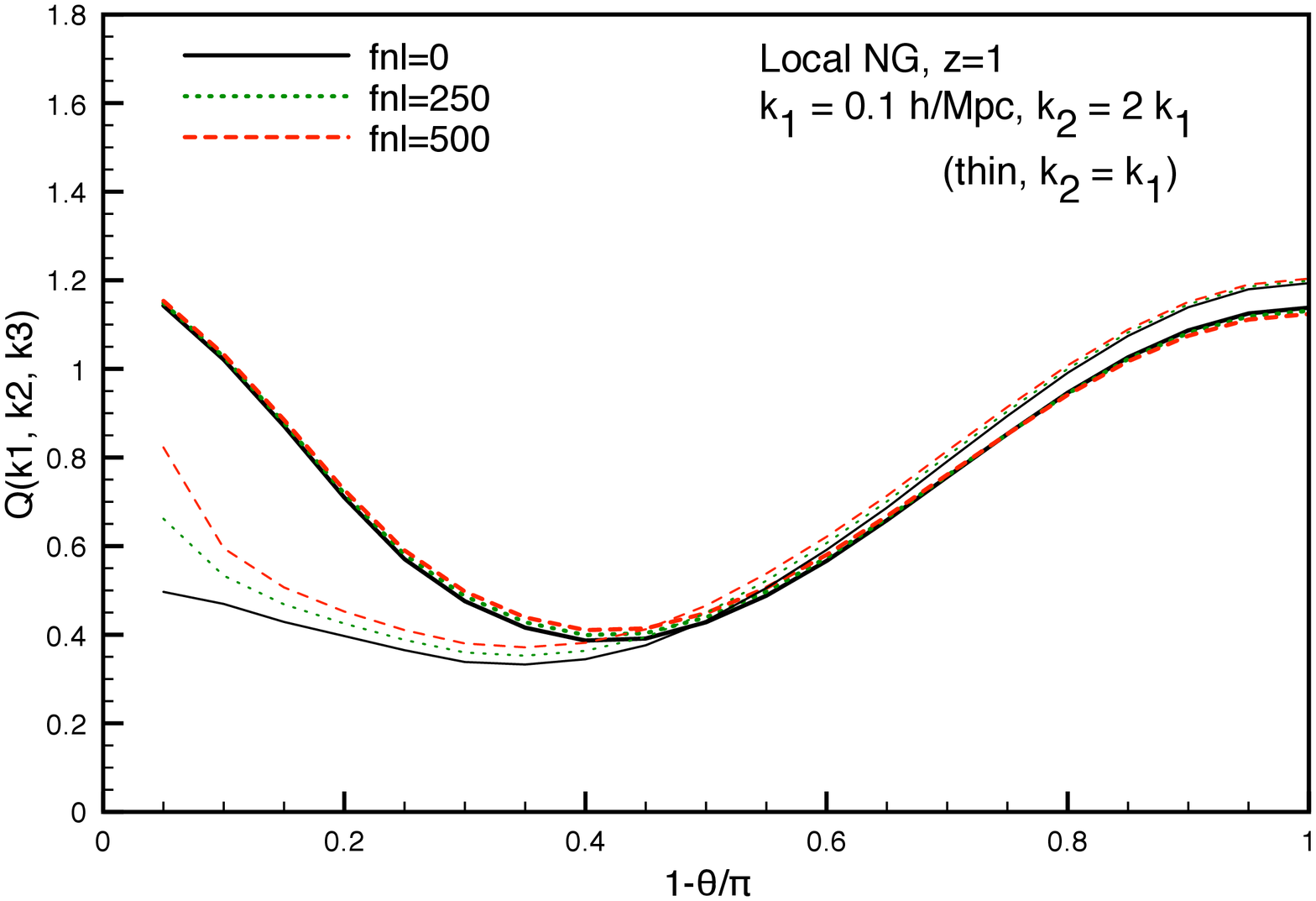} \includegraphics[width=3.in]{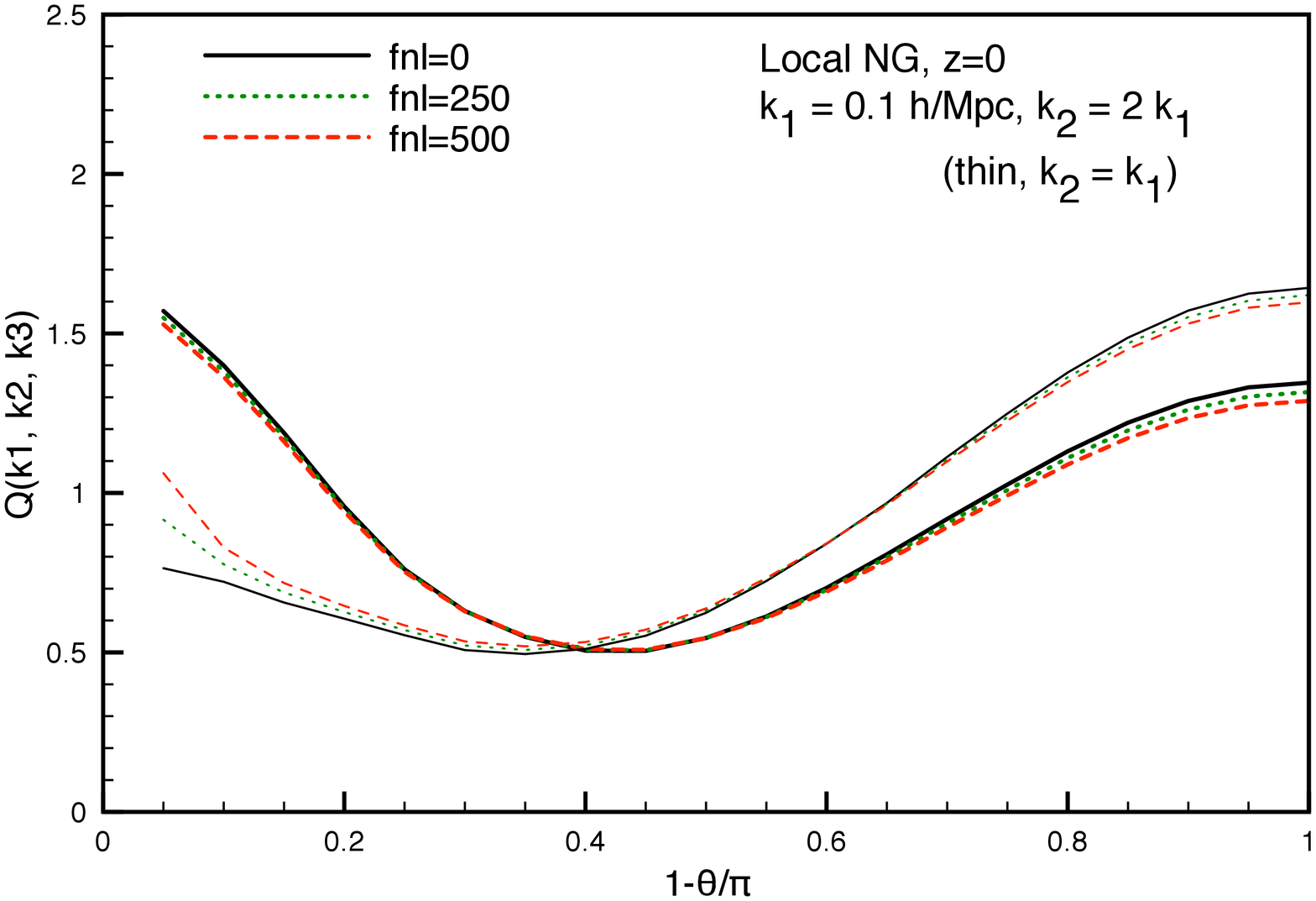}
        \caption{The reduced bispectrum for the local model of NG with fixed $k_{1}$ and ratio $k_{2}/k_{1}=2$ (thick lines) and  $k_{2}/k_{1}=1$ (thin lines). The left panels shows 
        the evaluation at $z=1$ while the right panels does it for $z=0$. In each case we plot $Q$ for non-Gaussian initial conditions with $f_{\rm NL}= 250$ (green-dotted) and $f_{\rm NL}= 500$ (red-dashed). 
        From top to bottom we plot $Q$ for $k=0.01,\, 0.05,$ and $0.1\,h$/Mpc respectively. The thin red-dashed and green-dotted lines in the upper panel show the divergent behaviour due to the non-Gaussian component near $\theta = \pi$ in the squeezed limit. The black-continuous thin line on the same panels represent the Gaussian component of the reduced bispectrum which approaches to zero as $\theta \rightarrow \pi$ in this configuration.  }      \label{Qlocalr2}
\end{figure}

The results of our evaluations for the local model of primordial NG can be seen in Figure \ref{Qlocalr2}. 
From the upper panels in Figure \ref{Qlocalr2}, which corresponds to $k=0.01\,h$/Mpc, we see that $Q$ is much more 
sensitive to the effect of the primordial  NG for small $k_{1}$. In this case we see that the effect of non-Gaussianities on $Q$ is almost independent of $\theta$ for $z=1$ and $z=0$. Nevertheless, for higher redshifts we can see clearly a maximization in the effect of the non Gaussianities in the squeezed limit which corresponds to $\theta \rightarrow \pi$. The effect is dramatically maximized when we consider the squeezed limit of the configuration $k_{2}/k_{1}=1$ as shown by the thin lines of this Figure. In this configuration, the condition $\theta \rightarrow \pi$ is equivalent to $k_3/k_{1} \ll 1$. We can understand the behaviour at this limit by studying the first terms in the perturbative expansion of $Q$. The first term containing the non-Gaussian contributions is $Q^{(0),\, \rm{NG}}(k_1, k_2, k_3)\equiv B^{(0),\, \rm{NG}}(k_1, k_2, k_3)/\Sigma^{0}(k_1, k_2, k_3)$. Going to the squeezed limit $k_{3}/k_{1}\ll 1$, this non-Gaussian contribution scales roughly as $Q^{(0),\, \rm{NG}}(k_1, k_1, k_3)\sim f_{\rm NL}/k_3^2$. This is why in the limit $\theta \rightarrow \pi$, the non-Gaussian component is huge, actually, it is divergent. For this reason we should not consider the evaluation at this extreme configuration as a precise prediction but as an indication of the behaviour at this limit. The exact scaling of $Q^{(0),\, \rm{NG}}$ depends also on the precise details of the transfer function which could alleviate up to some extent the divergent behaviour in this configuration for higher values of $k_1$ as suggested by the results in the lower panels of Figure \ref{Qlocalr2}. On the other hand, the tree level Gaussian component of the reduced bispectrum $Q^{(0)}(k_1, k_2, k_3)\equiv B^{tree}(k_1, k_2, k_3)/\Sigma^{0}(k_1, k_2, k_3)$, where $B^{tree}(k_1, k_2, k_3)$ is the Newtonian piece of the bispectrum, scales like $Q^{(0)}(k_1, k_1, k_3)\sim k_3/k_1$ in the squeezed limit and consequently it goes to zero for $\theta \rightarrow \pi$. This tendency can be seen in the black-thin lines in the upper panels of Figure \ref{Qlocalr2}. 
Finally, the impact of non Gaussianities for higher values of $k_{1}$ is much smaller and is almost independent of $f_{\rm NL}$ as can 
be seen in the bottom panel of Figure \ref{Qlocalr2} which corresponds to $k=0.1\, h$/Mpc.

\begin{figure}[h] 
\includegraphics[width=3.1in]{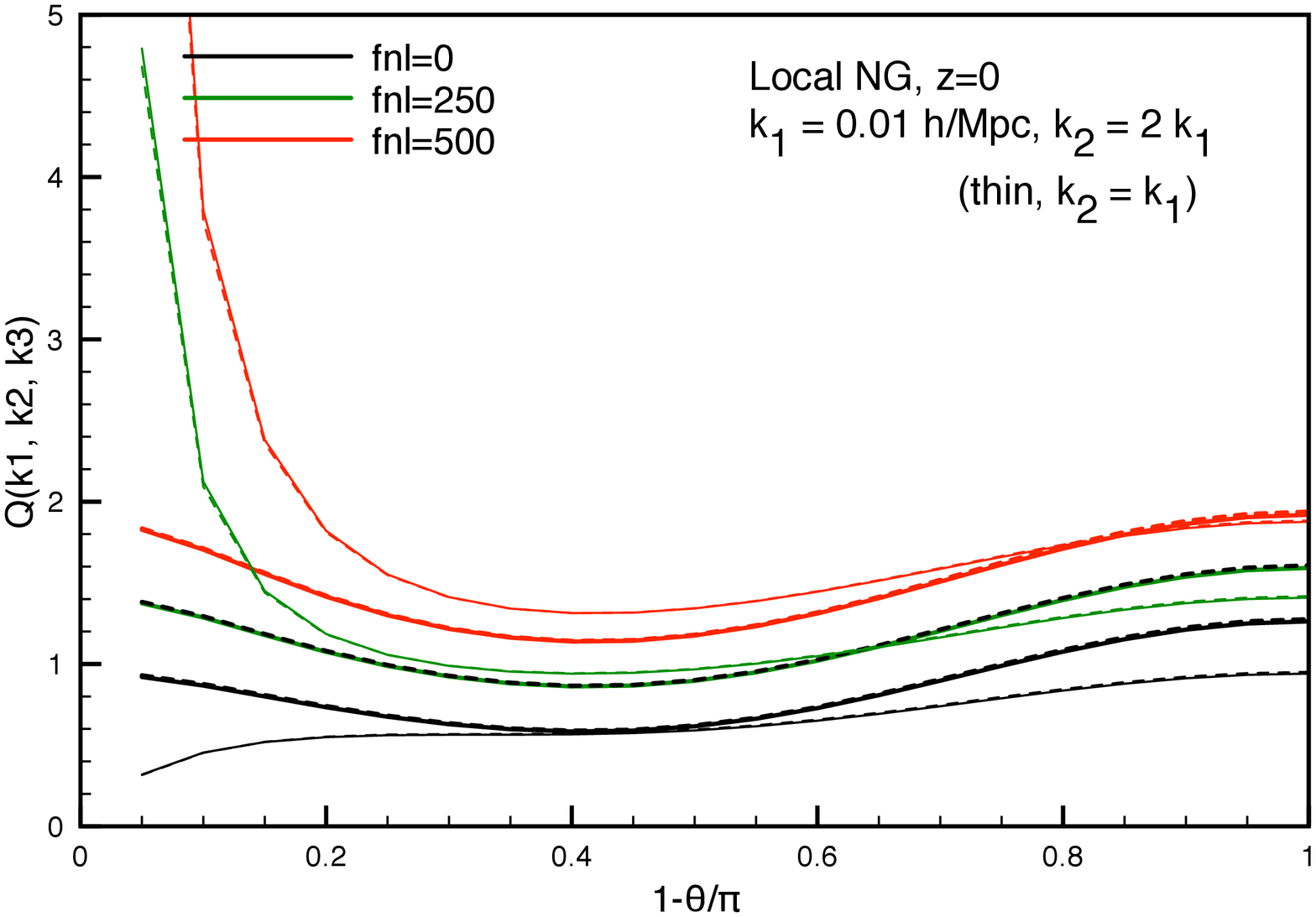} \includegraphics[width=3.1in]{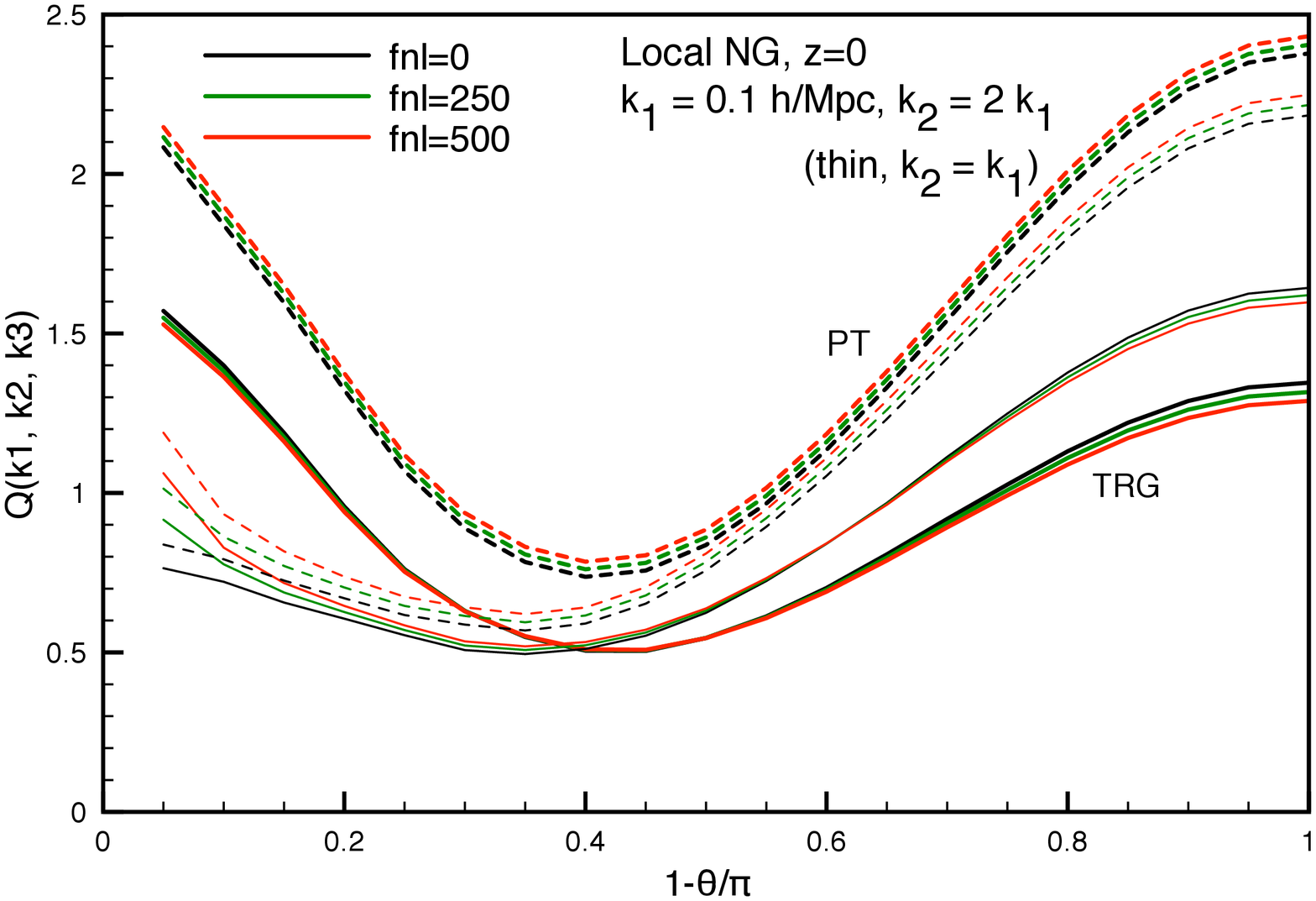} 
        \caption{Comparison of the tree level $Q^{(0)}$ and the TRG result for the reduced bispectrum. Continuous-thick lines are the TRG result and dashed-thick lines 
        represent the tree level evaluation at the configuration $k_2/k_1 =2$. Both evaluations are done at $z=0$ for different values of $f_{\rm NL}$ and for $k_{1}=0.01$h/Mpc (left), and $k_{1}=0.1$h/Mpc (right). Continuous and dashed thin lines represent the evaluation for $k_2/k_1 =1$.}      \label{Qlocalr2Q0}
\end{figure}

Figure \ref{Qlocalr2Q0} shows the comparison of the TRG evaluation with the perturbation theory 
(PT)  tree level reduced bispectrum $Q^{(0)}$ for several values the $f_{\rm NL}$ at redshift $z=0$. 
These plots are meant to help understanding the relevance of non linear effects at different scales and redshifts. It is noticeable that there is a 
remarkable change in the impact of non linear effects depending on the scales we are considering. For instance, in the left panel of figure \ref{Qlocalr2Q0}, 
we plot for $k_1 = 0.01\,h$/Mpc. At this scale, the $f_{\rm NL}$ dependence is strong, as we can see from the large separation in the solid lines which 
depends of the strength of primordial non-Gaussian effects. On the other hand, the effects from non-linearities quantified by the separation between dashed 
lines  and the solid lines is rather mild. In the right panel, we plot the same situation for $k_1= 0.1 \,h$/Mpc. At this scale, we observe 
that the situation is quite the opposite, the non-linear effects are stronger, while, the dependence on the primordial non-Gaussianities is much weaker. 

\subsection{Equilateral model}
\begin{figure}[h] 
\includegraphics[width=3.in]{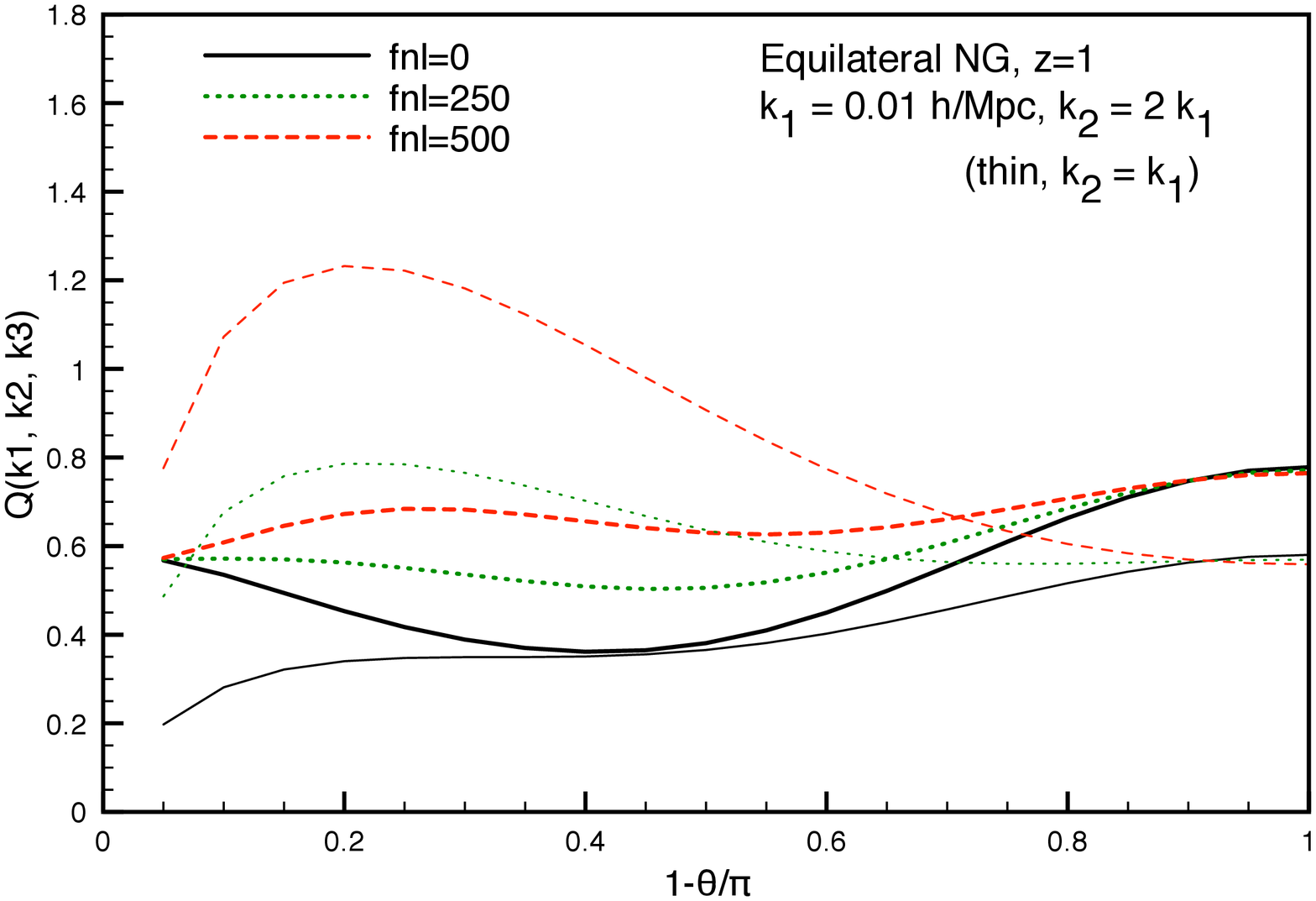} \includegraphics[width=3.in]{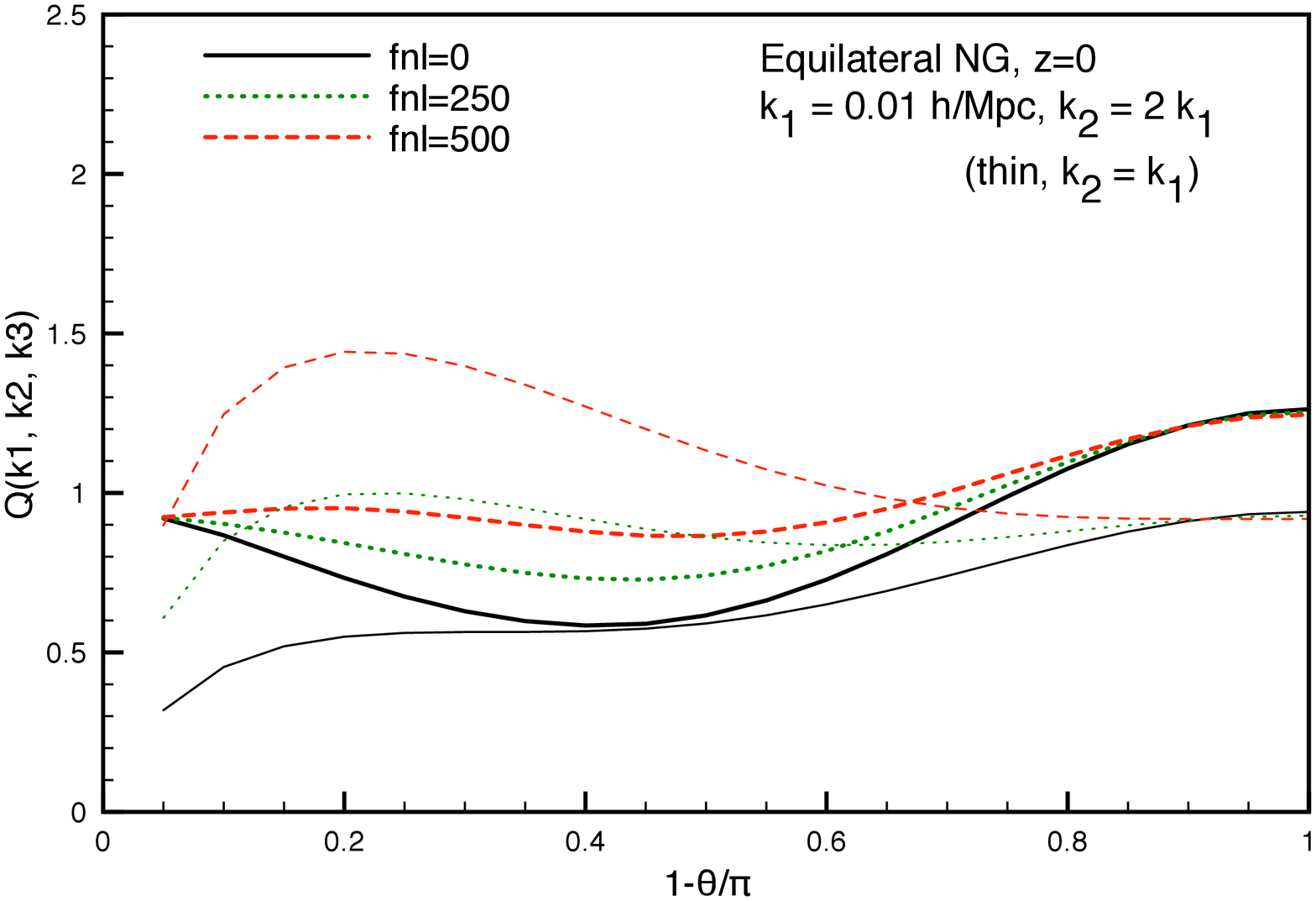} \\
\includegraphics[width=3.in]{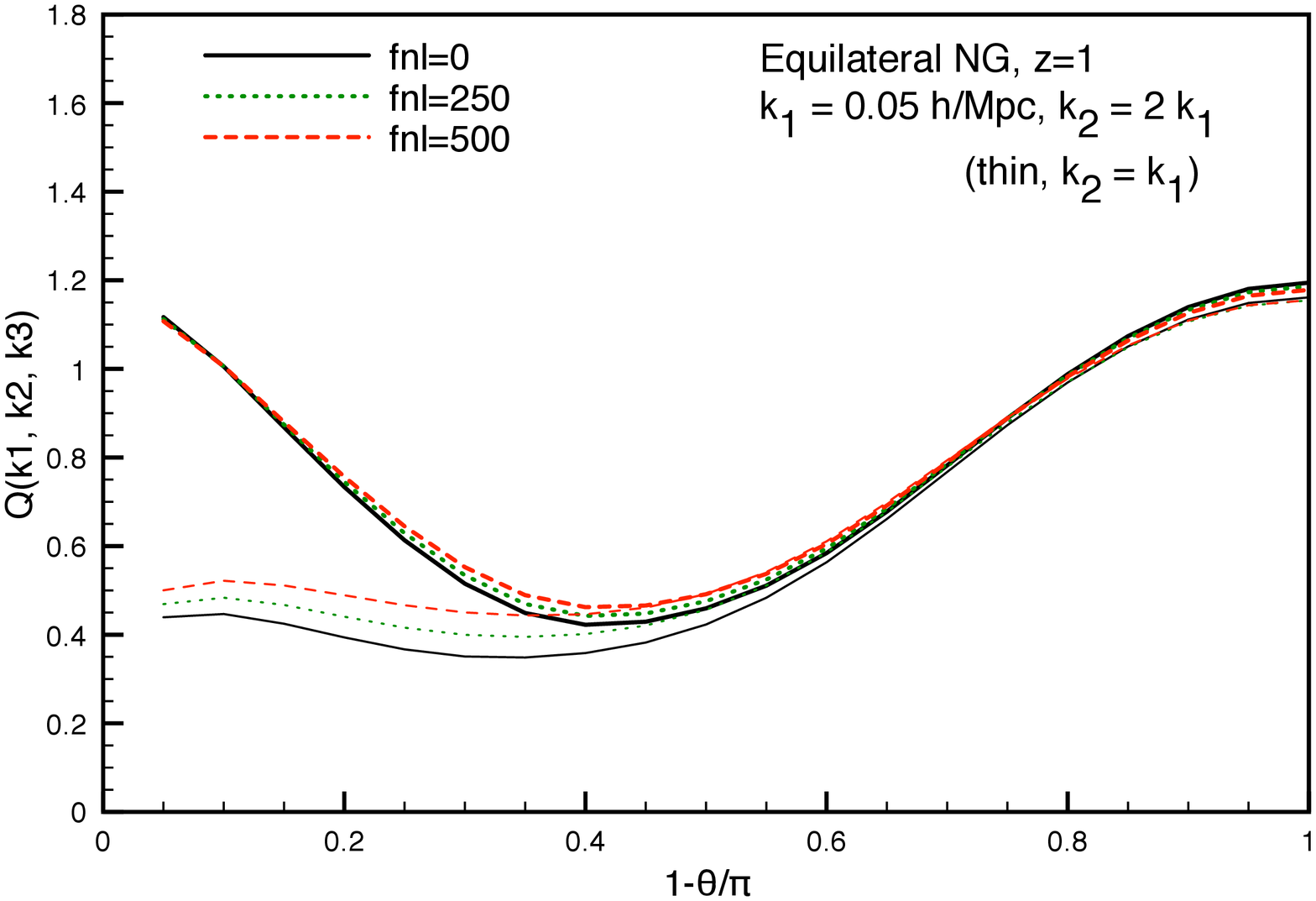} \includegraphics[width=3.in]{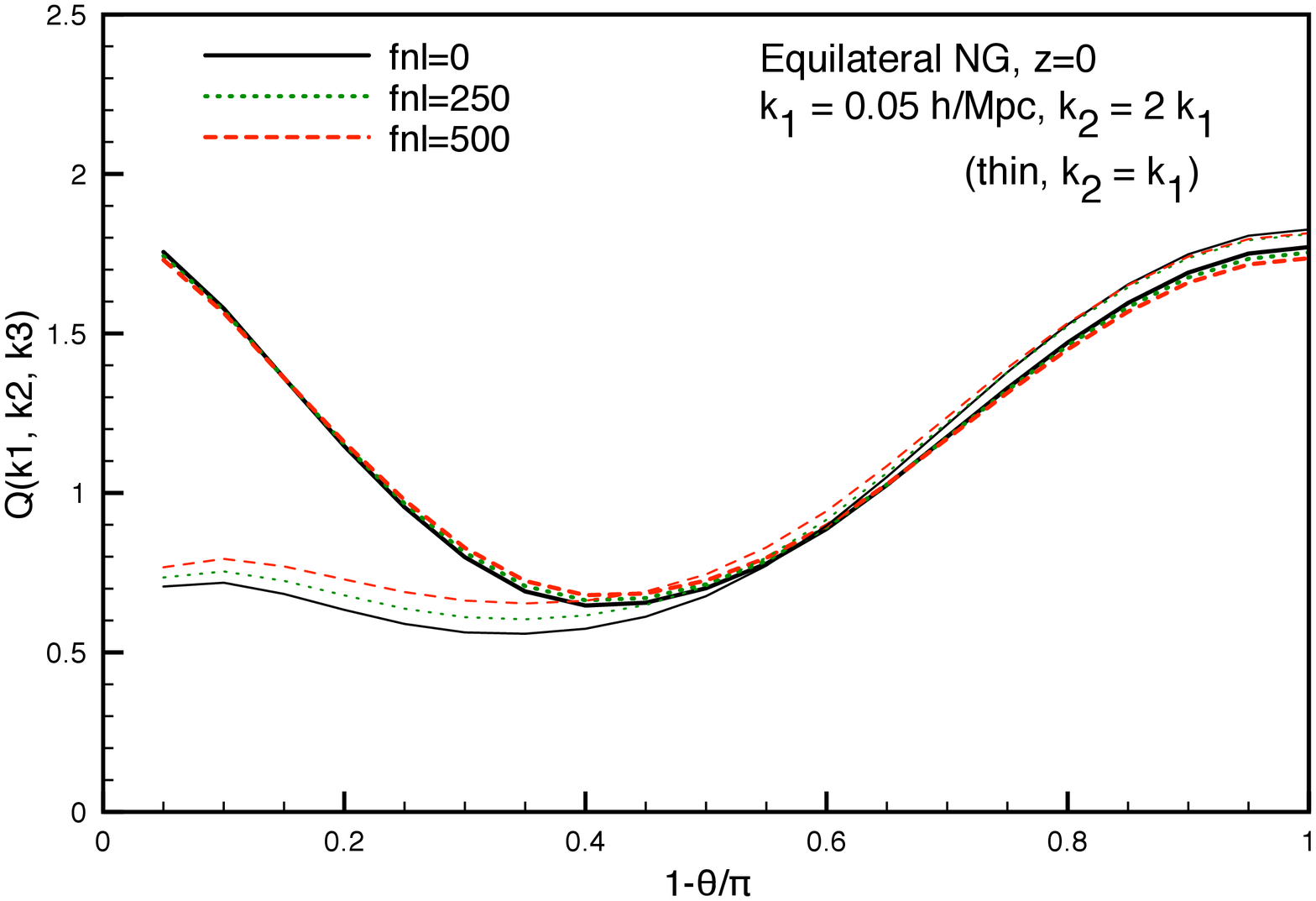} \\
\includegraphics[width=3.in]{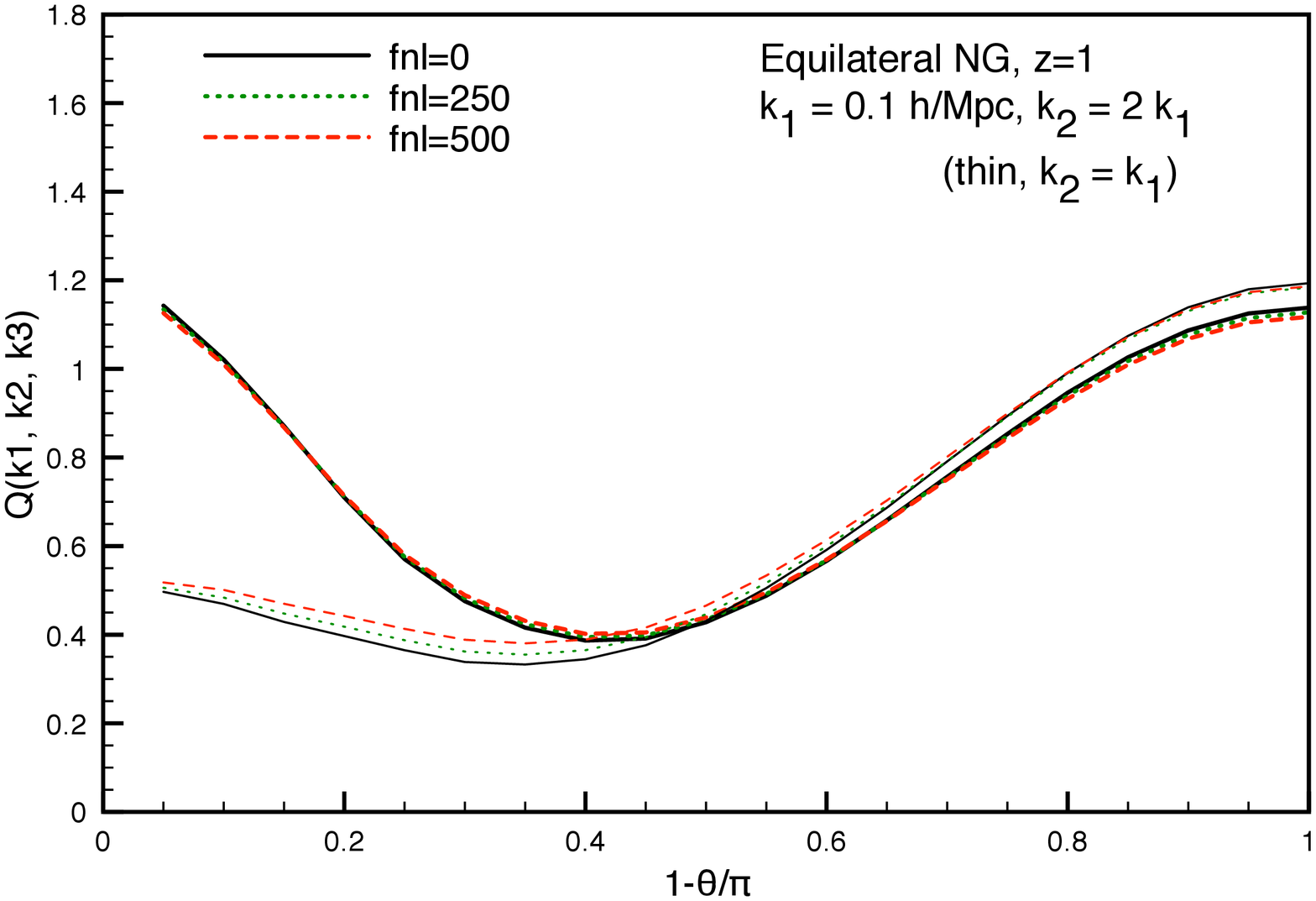} \includegraphics[width=3.in]{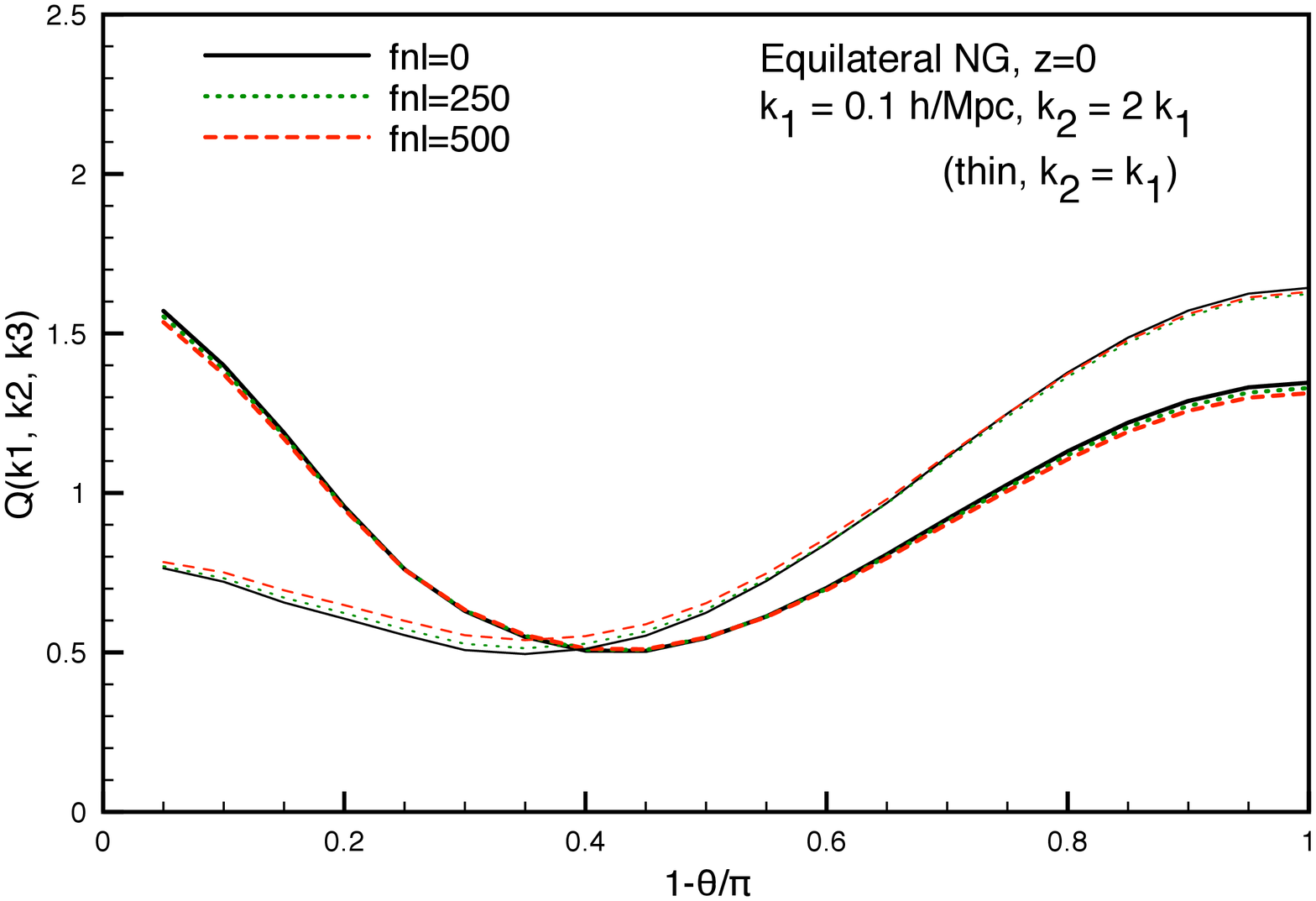}
\caption{The same as figure  8 for the equilateral model of primordial NG.}      \label{QEQr2}        
\end{figure}

The equilateral model exhibits a different pattern compared to the local model. This behaviour can be seen more clearly in the upper panel of figure \ref{QEQr2}. 
There, we see that $Q$ is enhanced by non-Gaussianities in a way that it is highly dependent on $\theta$. The maximum of the effect occurs at $\theta \sim 2 \pi /3$ which corresponds to equilateral configurations in this model and are responsible for the bump in this Figure near to  $\theta \sim 0.8 \pi $.  Again, as in the local model, the effect of non-Gaussianities is greatly enhanced in the configuration $k_2/k_1 =1$ while for higher $k_{1}$ the reduced bispectrum 
is much less sensitive to the effect of primordial NG regardless the value of the $f_{\rm NL}$ parameter.

\begin{figure}[h] 
\includegraphics[width=3.in]{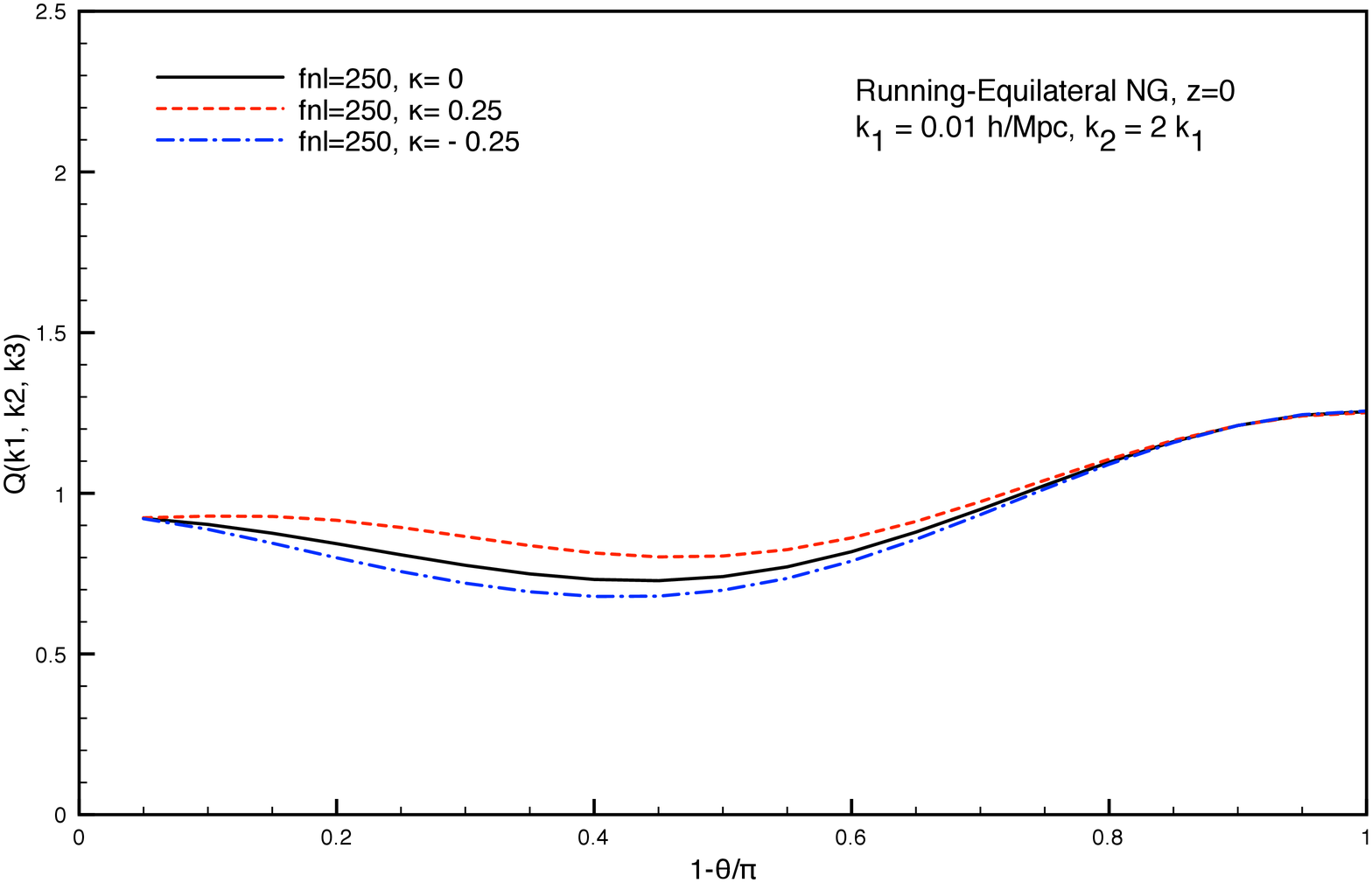} \includegraphics[width=3.in]{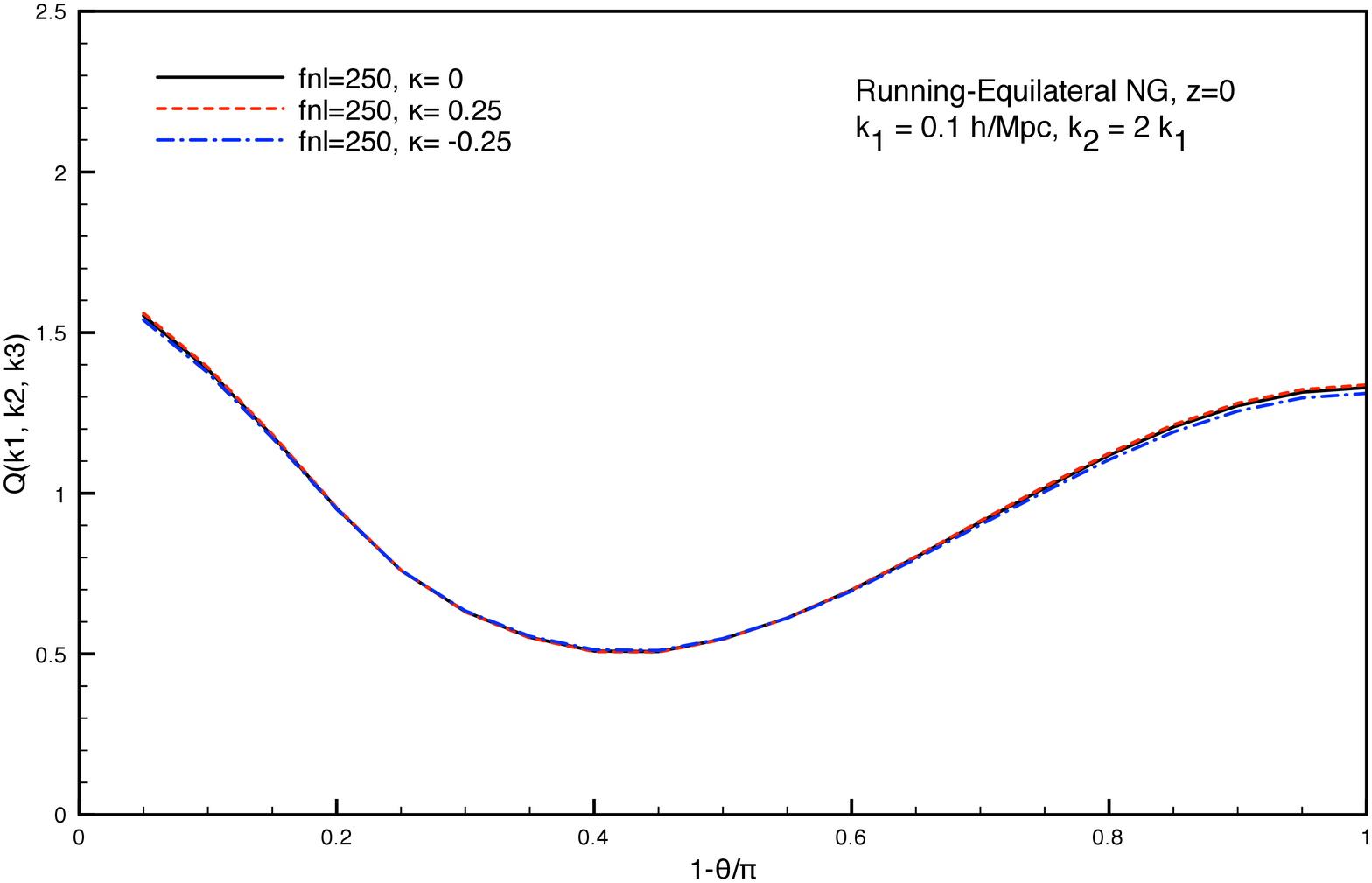} 
\caption{Reduced bispectrum for running $f_{\rm NL}$ with $f_{{\rm NL}, P}=250$ and $\kappa =0$ (solid, black line), $\kappa =0.25$ 
(dashed, red line) and  $\kappa =-0.25$ (dot-dashed, blue line).}\label{ReqQ}        
\end{figure}

The result of the evaluation of the running $f_{\rm NL}$ equilateral case is shown in figure \ref{ReqQ}. In the configuration $k_2/k_1=2$, we evaluate for $\kappa = 0, 0.25, -0.25$. For $k_1 = 0.01\, h$/Mpc and $z=1$ we find a 
remarkable enhancement (suppression) for positive (negative) values of $\kappa$. For $k_1 = 0.1\, h$/Mpc and $z=0$ the effect is rather unnoticeable. 

\subsection{Folded model}
\begin{figure}[h] 
\includegraphics[width=3.in]{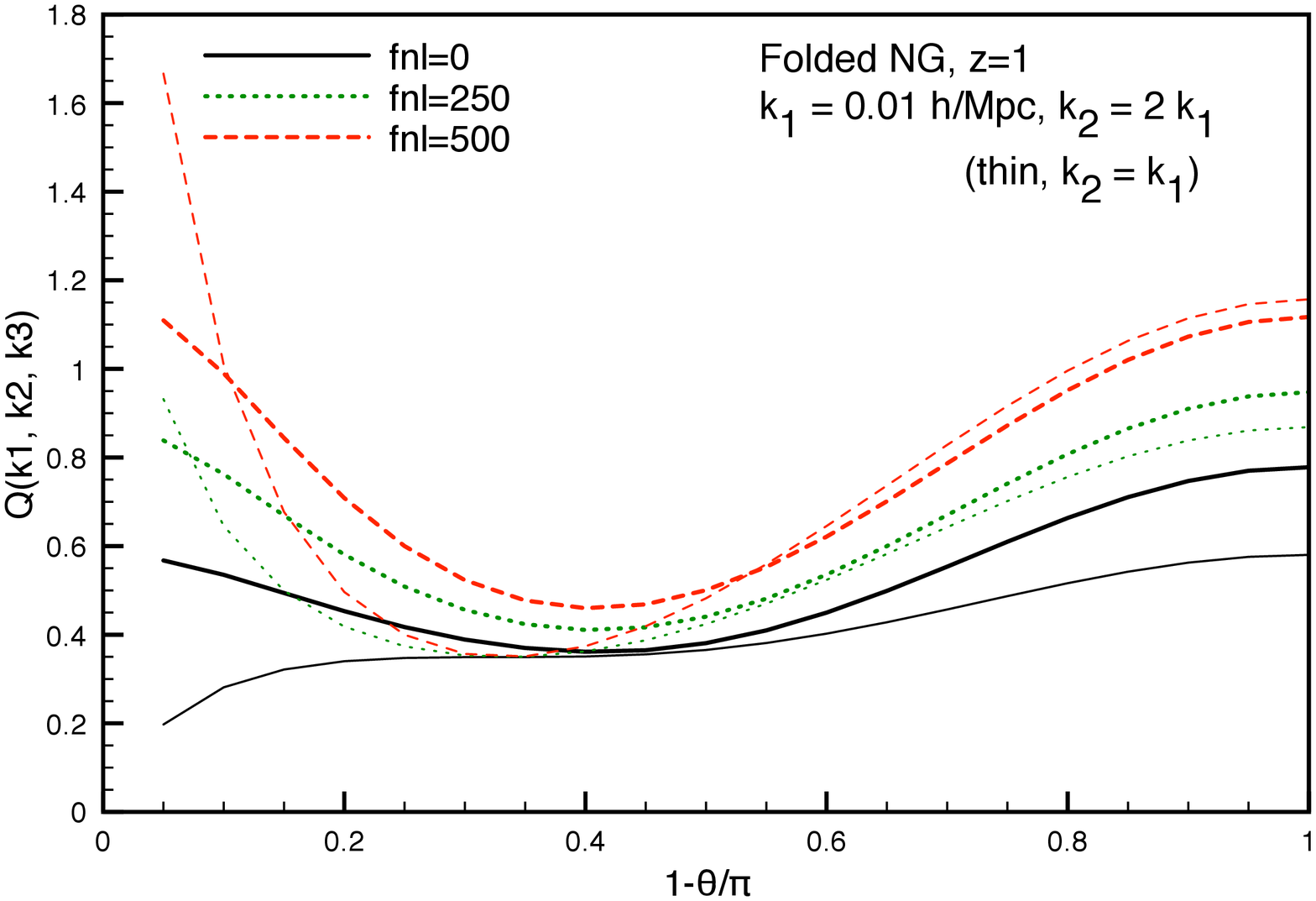} \includegraphics[width=3.in]{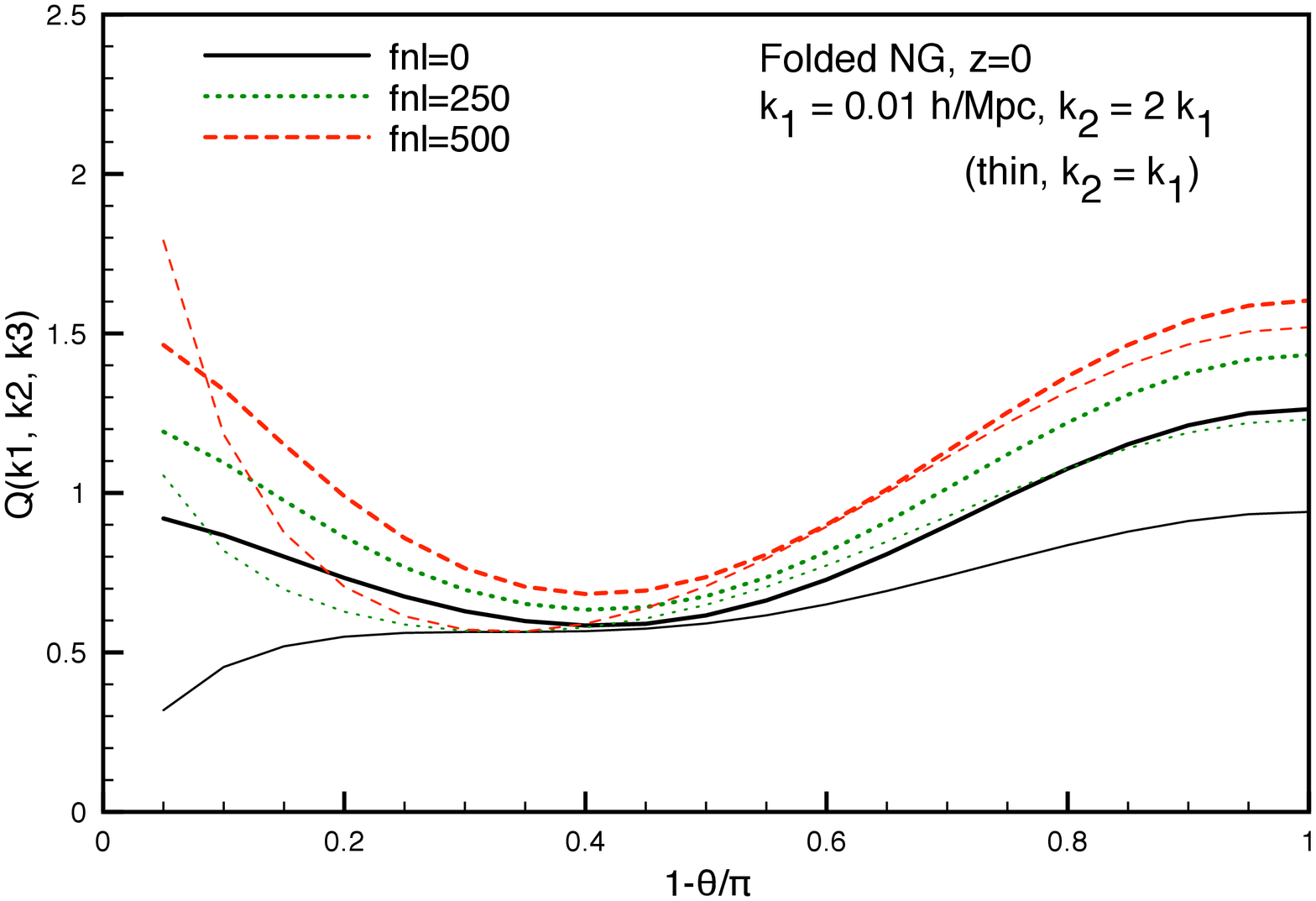}\\ 
\includegraphics[width=3.in]{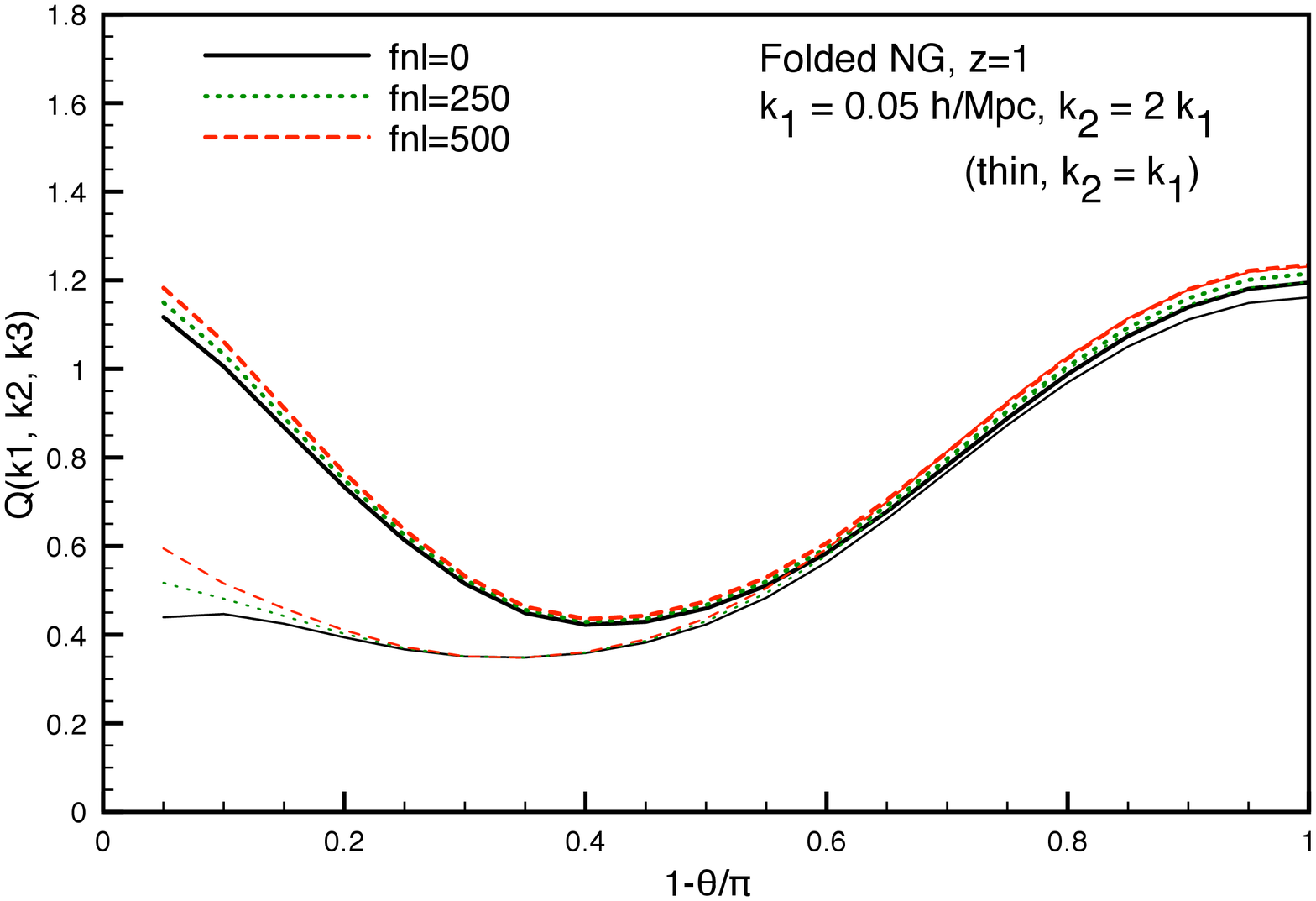} \includegraphics[width=3.in]{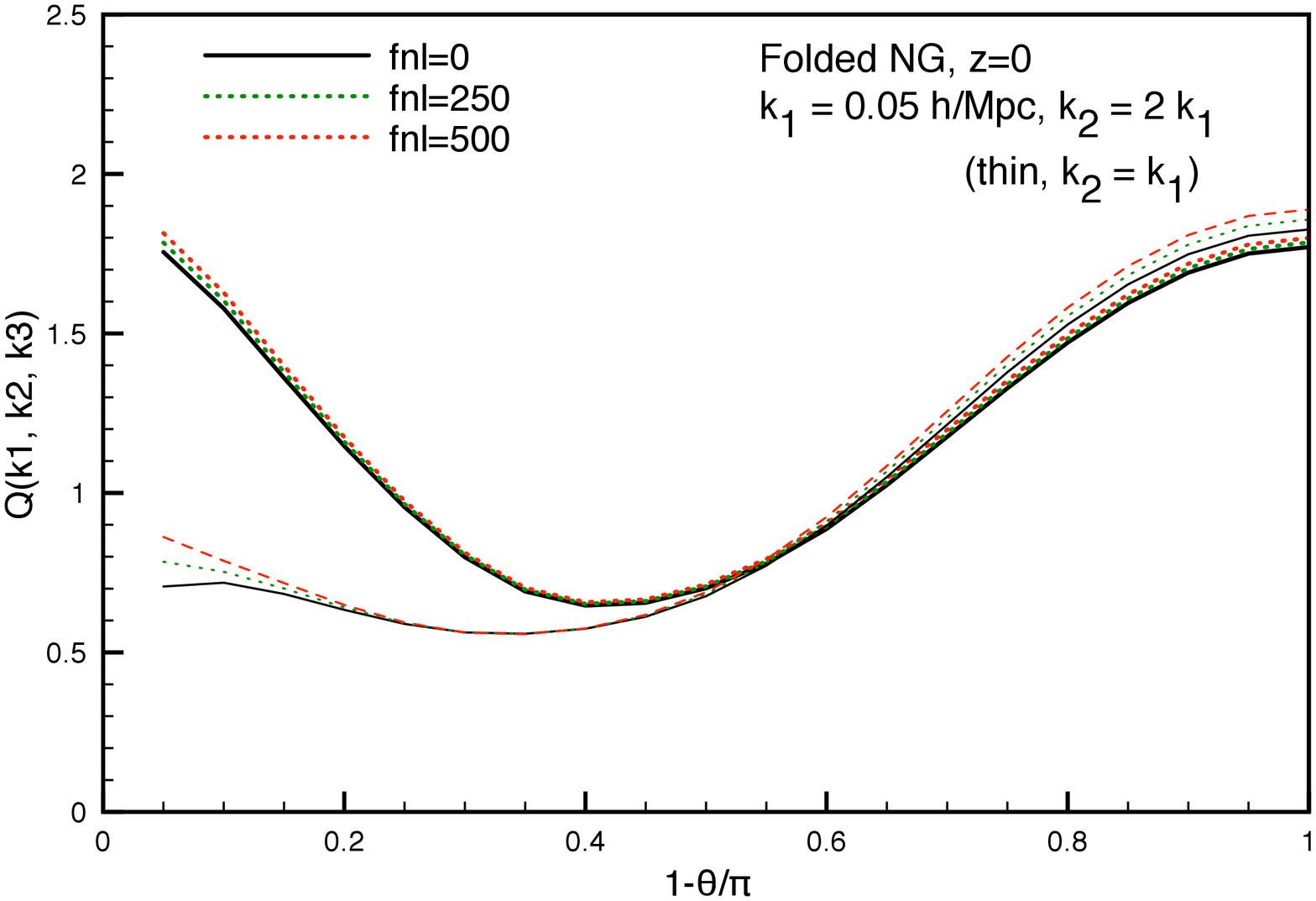}\\
\includegraphics[width=3.in]{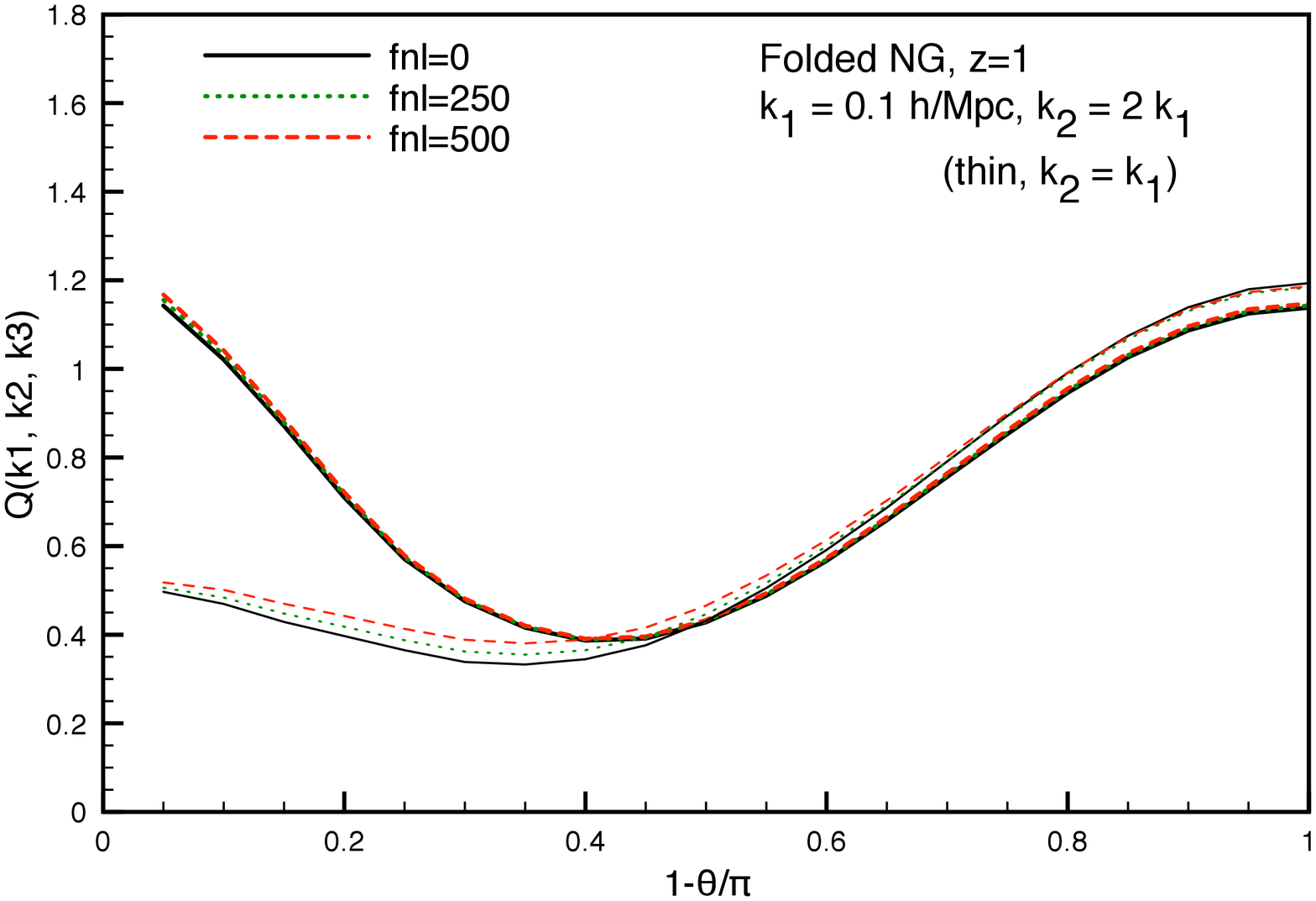} \includegraphics[width=3.in]{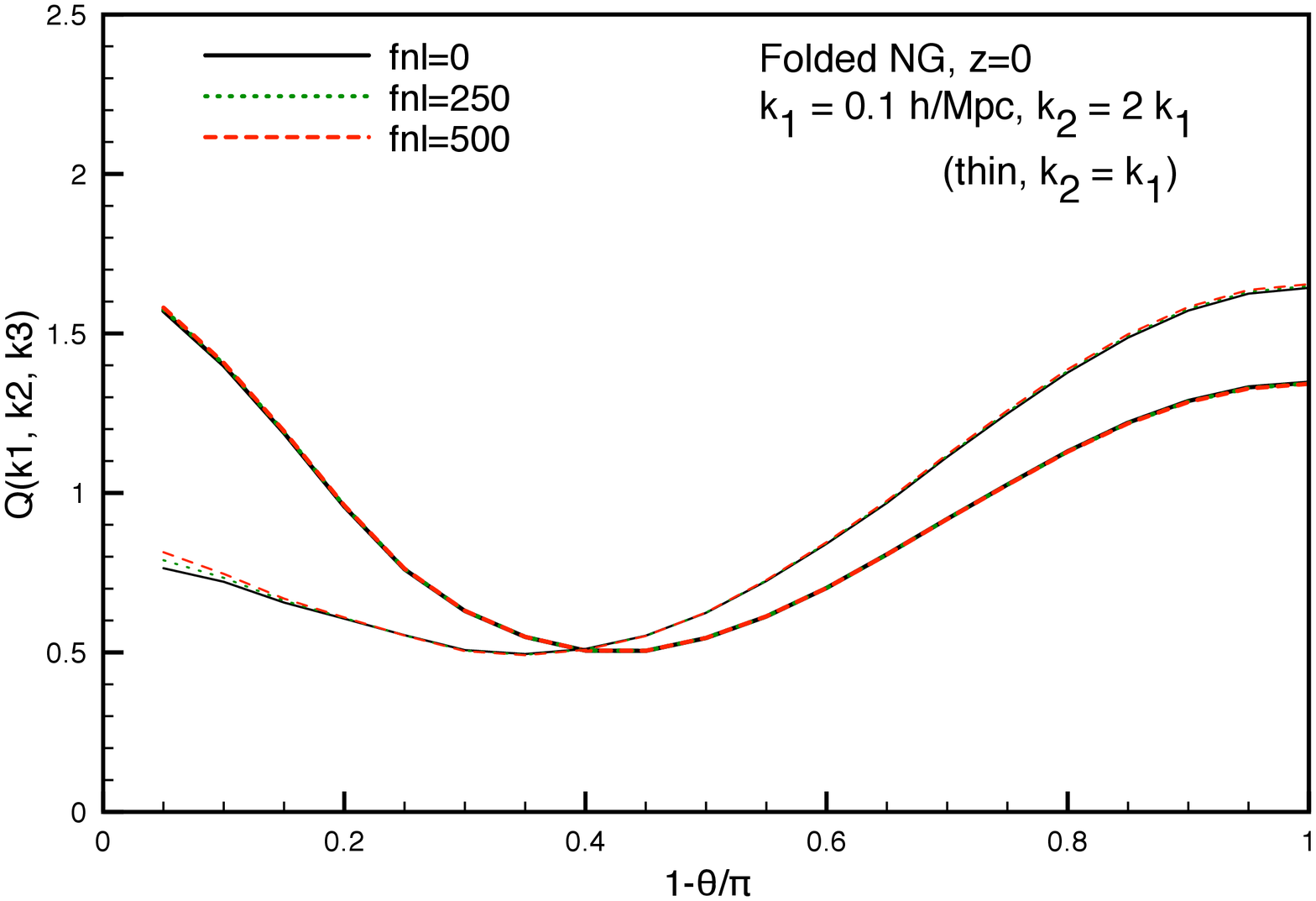}
\caption{The same as figure 8 for the folded model of primordial NG.}      \label{QFSr2}
\end{figure}

Contrary to the equilateral model, in the folded model, $Q$ is enhanced by the non-Gaussian contributions coming from the correlations among modes with almost collinear wave-vectors ($\theta = 0, \, \pi$). This effect is maximized in the limit $\theta \rightarrow \pi$ (as in the local model) which corresponds to a folded configuration of the momentum-space triangle when $k_2 \approx k_3 \approx  k_{1} /2$ (Figure \ref{shapes}). This effect can be seen in figure \ref{QFSr2}. As in the previus cases, the effect is maximized for small $k_{1}$ and for high 
redshift. For higher $k_{1}$ and lower redshift as in the bottom panel of figure \ref{QFSr2} we see that the effect of this model of non-Gaussianities is tiny, regardless the value of $f_{\rm NL}$.

\section{Conclusions and Discussion}
In this work we have studied several aspects related to the effects of primordial non-Gaussianities in gravitational clustering using the recently developed TRG method.
We have discussed different shapes of primordial NG and evaluated their impact in the non-linear evolution 
of the power-spectrum and (reduced) bispectrum of matter density perturbations. 

For the power-spectrum with some primordial NG present, we have compared our results to the ones  of standard one-loop perturbation theory and N-body data. 
In the range 
$0.1\lesssim k \,{\rm Mpc}/h \lesssim 0.2$, the one-loop and the TRG method reproduce pretty well the data from the N-body simulations of 
Ref.~\cite{Pillepich:2008ka} while, for $k \gtrsim 0.2\, h$/Mpc, the one-loop prediction   begins to show some inaccuracies which are within 
the $ \lesssim 2$ \% level. The non-linear growth of  the power-spectrum is more suppressed in the one-loop calculation than in the 
TRG calculation. The differences between the two approaches becomes more evident on smaller scales and lower redshifts, where the TRG 
offers a noticeable improvement with respect to the one-loop but still with a deviation from the N-body data. We expect that such deviations 
 can be reduced  if we  take into account  the ``vertex corrections" including the trispectrum in the system of equations as we discussed in the Introduction.
On general grounds, we expect that the effects arising from vertex corrections will become relevant for higher wavenumbers and that they will 
help to extend the range of validity of wavenumbers to larger values. We have also computed the
reduced bispectrum for various shapes of NG. As stressed in Ref. \cite{ks}, the reduced bispectrum 
is a powerful probe of NG. There the authors performed a Fisher matrix analysis to study the smallest 
value of the paramater $f_{\rm NL}$ measurable in high-redshift galaxy survays for local and equilateral shapes, after marginalizing 
over the bias parameters. At the tree-level the signal-to-noise ratio $(S/N)^2$ scales like (the sum over momenta forming a triangle of) $B^2(k_1,k_2,k_3)/P(k_1)P(k_2)P(k_3)$. 
For a local shape of NG, $k_1\ll k_2\sim k_3$, $B^2(k_1,k_2,k_3)/P(k_1)P(k_2)P(k_3)\sim P(k_1)$. 
Non-linearities enter in the determination of the maximum wavenumber $k_{\rm max}$ at which one computes $(S/N)\sim f^{\rm loc}_{\rm NL}\,
k_{\rm max}^{3/2}$. For the equilateral shape, $k_1\sim k_2\sim k_3$, and again at the tree level, the signal-to-noise ratio scales like $(S/N)\sim  f^{\rm eq}_{\rm NL}
\,k_{\rm max}^{5/2}\,P^{1/2}(k_{\rm max})\sim f^{\rm eq}_{\rm NL}\, k_{\rm max}$.  
Going beyond the linear order through the TRG method one should compute the full Fisher matrix and therefore $(\partial B(k_1,k_2,k_3)/\partial f_{\rm NL})^2$. In  this case 
the non-linear effects cannot be neglected. We leave this and other issues, such as the inclusion of the trispectrum,  for future work.

\section*{Acknowledgments}
We kindly acknowledge C. Porciani and T. Giannantonio for providing us with the data from N-body simulations with non-Gaussian 
initial conditions reported in \cite{Pillepich:2008ka} and \cite{Giannantonio:2009ak}. We also thank M. Grossi and L. Moscardini for useful discussions.
JPBA would like to acknowledge very kind hospitality and partial financial support from INFN - Sezione di Padova  during a visit and also from the Theory 
Division at CERN during early stages of this work. JPBA is supported by FAPESP under processes 2006/00622-8 and 2007/58202-7. ASI is acknowledged for partial financial support under 
contract I/016/07/0 ``COFIS" and through ASI/INAF Agreement I/072/09/0 for
the Planck LFI Activity of Phase E2. MP and AR acknowledge support by the EU Marie Curie Network "UniverseNet" (HPRN–CT–2006–035863).


\end{document}